  \def\om{\Omega} \def\ota{one-to-all }
\newcommand{\tXo}[1]{\tilde{X}_{A #1}}
\newcommand{\tXt}[1]{\tilde{X}_{B #1}}
\newcommand{\tPo}[1]{\tilde{Pn}_{A #1}}
\newcommand{\tPt}[1]{\tilde{P}_{B #1}}
\newcommand{\bXo}[1]{\bar{X}_{A #1}}
\newcommand{\bXt}[1]{\bar{X}_{B #1}}
\newcommand{\bPo}[1]{\bar{P}_{A #1}}
\newcommand{\bPt}[1]{\bar{P}_{B #1}}
\newcommand{\xo}[1]{x_{A #1}} \newcommand{\Xo}[1]{X_{A #1}} \newcommand{\xt}[1]{x_{B #1}}
\newcommand{\Xt}[1]{X_{B #1}} \newcommand{\po}[1]{p_{A #1}} 
\newcommand{\ptw}[1]{p_{B #1}}  \newcommand{\Mo}[1]{\tilde{M}_{ #1}}
\newcommand{\Mt}[1]{\tilde{M}_{ #1}} \newcommand{\Vo}[2]{V_{ #1 #2}\left( \xo{#1}-\xo{#2} \right)}
\newcommand{\Vt}[2]{V_{ #1 #2}\left( \xt{#1}-\xt{#2} \right)}

\def\be{\begin{equation}} \def\ee{\end{equation}} \def\te{\end{equation}} \def\bea{\begin{eqnarray}}

\def\og{\lambda}
\def\fg{g}
\def\fgm{\mathbf{g}}
\def\tg{G}
\def\tgm{\mathbf{G}}
\def\sg{\tilde{G}}

\def\n{n}

\def\nA{A}

\def\bfX{\mathbf{X}}
\def\nX{X}



\documentclass[preprint]{elsarticle}
\usepackage{amsmath,amssymb}
\usepackage{xcolor}
\usepackage{pdfcomment}
\usepackage{soul}
\usepackage{graphicx,color}
\usepackage{amssymb,amsmath,amsthm,graphicx,amscd}
\usepackage[mathscr]{eucal}
\usepackage{enumerate,color,verbatim,multirow,comment}
\usepackage{bm}
\usepackage{float}

\journal{Physica A}

\begin{document}

\baselineskip=18pt
\numberwithin{equation}{section}

\allowdisplaybreaks

\begin{frontmatter}

\title{Macroscopic Quantum Phenomena  from the Coupling Pattern and Entanglement Structure Perspective}

\author{C. H. Chou}
\ead{chouch@mail.ncku.edu.tw}
\address{Department of Physics, National Cheng Kung University, Tainan, Taiwan}
\author{Yi\u{g}it Suba\c{s}{\i} and B.~L.~Hu}
\ead{ysubasi@umd.edu, blhu@umd.edu}
\address{Joint Quantum Institute and Maryland Center for Fundamental Physics, \\ University of Maryland, College Park, Maryland 20742}


\begin{abstract}

Entanglement being a uniquely quantum attribute, to gain a better understanding of the nature of macroscopic quantum phenomena we explore in this paper ways to qualify and quantify the quantum entanglement $E(M)$ between two macroscopic objects by way of model studies. Knowing that a macroscopic object is a composite, how does one determine $E(M)$ in terms of the entanglements between its micro-constituents $E(\mu)$? We assert that the notion of `levels of structure', the coupling strength between constituents in different levels, and the use of collective variables to capture the contribution of each level are all pertinent considerations. To understand how the {coupling pattern amongst the constituents} of the two macro objects enters into the picture, we consider two types of coupling, each constituent particle is coupled to only one other particle (1-to-1) versus coupled to all particles (1-to-all). In the 1-1 case with pairwise interactions of equal strength, the entanglement is independent of the number of constituent particles $N$ in the macroscopic object. In the 1-to-all case the relative coordinates are decoupled and the center of mass (CoM) coupling scales with $N$. Here we expect the entanglement between the CoM variables to increase with increasing size of the macroscopic objects and survive at higher temperatures. For a quantum many body system containing $N= 2^\n$ constituents, we provide a proof of the conditions whereby the CoM variable decouples, a cause for the special role the CoM variable plays in the entanglement between two such macroscopic objects. Similar qualitative behavior is found when the couplings between the constituents of the macroscopic objects are statistically independent Gaussian random variables. In another model study we analyzed the entanglement pattern of 4 coupled oscillators in two pairs, representing the two objects A and B (or two adjacent levels of structure), each with two constituents. By assigning different coupling strengths we can investigate the interplay of inter-level entanglement with intra- level interactions representing tight or weak binding of the constituents within one level of structure. From the entanglement dynamics of the 4-oscillator system with varying coupling strength we see the entanglement between constituents meeting sudden death while the CoM variables may sustain over longer times. This offers another way to determine under what conditions quantum entanglement between macroscopic objects can persist. We explore the implications of our findings for quantifying the entanglement measure for macroscopic objects which is different from that of the entanglement entropy dependent on the area of a partition for a quantum many body system. Finally we mention how the model studies here bear on quantifying the entanglement in physical systems such of nucleons in a nucleus, layers of graphene and between DNA macromolecules.

\end{abstract}


\end{frontmatter}



\tableofcontents

\newpage

\section{Introduction}

Macroscopic quantum phenomena (MQP) refer to quantum features in objects of `large' sizes, systems with many components  or degrees of freedom, organized in some ways where they can be  identified as macroscopic objects. What makes MQP interesting is because in the traditional view `quantum' belongs to the microscopic domain, in the sense that only 'small' objects necessitate a quantum description, while classical mechanics,  a limiting case of quantum mechanics, is sufficient for the description of the macro world. In the face of new challenges from macroscopic quantum phenomena, viz, quantum features may show up even at macroscopic scales, this common belief now requires a much closer scrutiny, involving possible reformulations and/or re-interpretations.  In addition to the worthy effort invested in the last two decades to gain a better understanding of the quantum-classical relation, as was the aim of quantum decoherence studies, and what could be viewed as characterizing the `quantumness' of a system, such as the existence of quantum entanglement in the system,  we note that faced with the challenge of MQP, even a naive and seemingly unequivocal  notion, like what is meant by `macroscopic', need be re-considered.

This emerging field is ushered  in by several categories of definitive
experiments. A common example of MQP is superconductivity, where the
Cooper pairs can extend to very large scales compared to interatomic
distances, and Bose-Einstein condensate (BEC), where a large number N
of atoms occupy the same quantum state, the N-body ground state. Other
important examples are in  nanoelectromechanical devices \cite{nem},
where the center of mass of a macroscopic 
object, the cantilever, is seen to obey a quantum mechanical equation of motion. Experimental proposals to detect the superposition between a mirror and the quantum field, and between two mirrors, have been proposed \cite{Marshall,EntSQL} while the interference pattern formed when a large object composed of $C^{60}$ molecules passing through two slits have been observed \cite{Arndt}. Likewise for experiments in quantum optomechanics, see e.g., \cite{Aspelmeyer,Eisert}.

By contrast, this new field, which is rich in open  issues at the foundation of quantum and statistical physics,  remains little explored theoretically (with the important exception of the  work of Leggett \cite{Leggett}). Our attitude on this issue is not to give up quantum mechanics at a certain meso or macro scale, but to examine carefully the conditions whereby quantum features may survive at such scales, or if they don't, we want to know how and why? There is also a subtle difference between the term `phenomena' in MQP and macroscopic quantum mechanics (MQM) \cite{ChenMQP}. As we see it,  MQP explores the quantum features in macroscopic systems based on quantum mechanics whereas MQM conjures up the possibility that quantum mechanics as a well-tested theory for microscopic objects may either yield different results, or need be replaced by a different theory, at the macroscopic level.

In two recent essays \cite{MQP1,MQP2} we have suggested two pathways toward understanding MQP. The first concerns what macroscopic means -- large size? large number of components? For this we used the $O(N)$ quantum mechanical model to examine what exactly does the leading order and next-to-leading order large N behaviour of this theory say -- At the large N limit does the system become classical, as often conjured? We find even at the mean field level when a Gaussian system obeys an equation with the same form as the classical equation of motion (cf Ehrenfest theorem), the system is not classical but remains quantum in nature.  We reckon that there is no a priori good reason why quantum phenomena in macroscopic objects cannot exist. The second pathway explores how quantum correlations and fluctuations impact on MQP. Examples we used to examine this aspect include the running of coupling constants with energy and scale, the critical behavior of a system near phase transition, with the help of two-particle-irreducible (2PI) effective action and effective dimensional reduction in the  infrared regime, in the spirit of Kaluza-Klein theories. This perspective offers a dynamics-dependent  and an interaction-sensitive definition of ``macroscopia''.

This paper explores a third avenue, from the quantum entanglement
perspective. Since, according to Schr\"odinger \cite{Schrodinger} this
is an  exclusively quantum feature, it should be a good measure of
quantumness in a system. Quantum entanglement has been applied
extensively to qualify  quantum phase transition (at zero temperature)
\cite{Sachdev,Zanardi}. It is also claimed to persist at high
temperatures \cite{VedralRMP} and large scales \cite{LutzChain} under
specific conditions and special domains.
Traditionally one expects classical behaviour at high
temperatures and large scales, not quantum.

To explore the entanglement between macroscopic objects,  such as had been suggested in terms of mirror superposition
\cite{Marshall,Bouwmeester}, one needs to first ask two questions:  1)
What constitutes a macro object? We mentioned large size, large
number. But how they are put together certainly makes a difference.
What about the degree of complexity of its constituents? What if the
constituents are non-interacting versus interacting? Weakly
interacting versus strongly interacting? 2) Which parties are being
entangled? Rather, as in the quantum world in principle everything may be
entangled with everything else, the essential question is, are there `representative'  physical variables which capture the essential features and correctly depict the dynamics of a macroscopic object as a distinct entity?
For example, why is it that an ostensibly macroscopic object such as a
cantilever should follow a quantum equation of motion? -- What is the
physical variable of a macroscopic object interacting with an
environment which behaves quantum mechanically?  More specifically, why is it that its center of mass variable does and not other variables, at least not ostensibly?

This ``center-of-mass axiom'' is implicitly assumed in many descriptions of MQP but rarely justified. The conditions upon which this can be justified are explored in \cite{CHY} with the derivation of a master equation for a system of N coupled quantum oscillators (NOS) weakly interacting with a finite temperature bath made of n quantum oscillators, in the form of the HPZ master equation \cite{HPZ}, extending earlier work for 2 coupled oscillators \cite{ChouYuHu}. (A mathematically more vigorous and complete treatment of NOS system is given in \cite{FlemingNHO}.)

The above dissection immediately reveals the importance of how the basic constituents of the objects interact and are organized in addition to the simple characterization of micro-meso-macro by their scales and total masses.  The aim of this paper is to find out a way,  amongst presumably many,  to quantify quantum entanglement between macroscopic objects. To do so we need to first understand how different levels of structure in a macroscopic object contribute to, or partake of, the entanglement with another object of similar structures.  For this purpose in Sec. 2 we describe the level of structure concept (see, e.g., \cite{E/QG} in a composite body, the importance of choosing the proper collective variables,  of which the center of mass is the most directly representative, to capture its behavior and dynamics. We revisit the center of mass axiom shown in an earlier paper and apply it to address the present issue.  Specifically, for a system composed of N constituents (particles or components), we want to see what kind of coupling or under what conditions the entanglement would scale with N.  In Sec. 3 we consider two types  of coupling configurations, first (Fig. 1a),  when any one constituent of Object A is coupled to one specific constituent of Object B (called 1-1 or pairwise interaction). The second type  (Fig 1b),  is when each constituent of Object A is coupled to all the constituents of Object B (called 1-to-all interaction). We will show that the latter type is when MQP would appear, as this interaction configuration scales with N. Next, we provide another proof of the conditions whereby the CoM variable decouples in a quantum many body system containing $N= 2^\n$ constituents, and with this, the special role the CoM  variable plays in the   entanglement between the two such macroscopic objects. In Sec. 4 we look more closely into the specific case of 4 coupled oscillators in two pairs,  representing the two objects A (constituents 1, 2) and B (constituents 3, 4). We assign different coupling strengths: $\alpha$ as that between the two (intra-level) constituents 1-2 or 3-4  in each object, $\beta$ as the (inter-level) coupling strength between 1-3 (and also 2-4) (parallel), and $\gamma$ between 1-4 (and also 2-3) (cross).  Considering $\alpha$ to be large compared to $\beta$ or $\gamma$ refers to strong intra- level interactions or tight binding of constituents within one object. This model can show the difference between the cases studied in Sec. 3, namely, the pairwise (1-3, 2-4) interaction versus the one to all (1-3 and 1-4) interaction.  As in Sec. 3 we see that it is only in the one-to-all case that the entanglement scales with N, assuming that entanglement increases with coupling strength.   
We then show the entanglement dynamics of the 4-oscillator system with varying coupling strength, while distinguishing three cases of entanglement: The case between the center of mass variables (CoM) of A and B,  the case between the CoM variable of one object A and one constituent particle of B, and the case between one constituent particle of A and one of B. We see some pariwise entanglement meets sudden death while others sustain over long time ranges. This offers another way to determine under what conditions quantum entanglement between macroscopic objects can persist.  Finally in Sec. 5 we summarize the key findings of this paper, compare with earlier results,  and discuss their implications for the existence of MQP. The model studies here can in principle be applied to a number of realistic experiments in cold atom and condensed matter systems.

\section{Levels of structure and the special role of collective variables}

`Macroscopic' conveys a sense of being `large', but what exactly does `largeness' mean?  Do all the
basic constituents of a large object contribute equally towards its quantum feature? (This point is highlighted
in footnote 2 of \cite{CHY}.) In some cases we may actually know what the basic constituents are and how they are organized. {} A $C^{60}$ molecule is made of carbon atoms, each atom is made of nuclei and electrons, each nucleus contains a
certain number of protons and neutrons, each of them in turn is made up of quarks and gluons. Are we to simply
count the number of quarks /gluons or protons /neutrons when we say an object is macroscopic? Obviously the tight
binding of them to form a nucleus enters into our consideration when we treat the nucleus as a unit which
maintains its own more or less distinct identity, features and dynamics. Thus when one talks about the mesoscopic or
macroscopic behavior of an object one needs to specify which level of structure is of special interest, and how
important each level contributes to these characteristics. The coupling strength between constituents at each
level of structure (e.g., inter-atomic) compared to that structure's coupling with the adjacent and remaining
levels (which can be treated as an environment to this specific level of structure in an effective theory
description, and its influence on it represented as some kind of noise  \cite{CalHuEFT,CalHu2000}) will determine the relative weight of each level of structure's partaking of the macroscopic object's overall quantum behavior. Often the best description of the behavior and dynamics of a particular level of structure is given by an effective theory for the judiciously chosen ``collective variables".

\subsection{Choose the right collective variables before considering their quantum behavior}

Same consideration should enter when one looks for the ``quantumness" of an object, be it of meso or
macro scale. One can quantize any linear system of whatever size, even macroscopic objects,  such as sound waves
from their vibrations. Giving it a name which ends with an ``on'' such as phonon and crowning it into a quantum
variable is almost frivolous compared to the task of identifying the correct level of structure and finding the
underlying constituents -- the atoms in a lattice in this example, and their interactions.   Constructing the
relevant collective variables which best capture the salient physics of interest should come before one considers
their quantum features. Thus, viewed in this perspective in terms of collective variables,  we see that  quantum
features need not be restricted to microscopic objects. In fact `micro' is ultimately also a relative concept as
new  ``elementary" particles are discovered which make up the once regarded `micro' objects.

We illustrate this idea first with a discussion of the relevance of the center of mass variable in capturing
the quantum features of a macroscopic object, then in the following sections, with a description of  quantum entanglement between two macroscopic objects.

\subsection{The quantum and macroscopic significance of center of mass variable}

We can ask the question: what are  the conditions upon which the mechanical and statistical
mechanical properties of a macroscopic object can be described adequately in terms mainly of its center-of-mass
(CoM) variable kinematics and dynamics,  as captured by a master equation (for the reduced density matrix,  with the
environmental variables integrated out). This is an implicit assumption made in many MQP investigations, namely,
that the quantum mechanical behavior of a macroscopic object, like the nanoelectromechanical oscillator
\cite{MQPnem,nem}, a mirror \cite{Marshall}, or a $C^{60}$ molecule \cite{Arndt}, placed in interaction with an
environment -- behavior such as quantum decoherence, fluctuations, dissipation and entanglement--  can be
captured adequately by its CoM behavior. For convenience we refer to this as the `CoM axiom'. This assertion is
intuitively reasonable, as one might expect it to be true from normal- mode decompositions familiar in classical
mechanics, but when  particles (modeled by NHO) interact with each other (such as in a quantum bound state
problem) in addition to interacting with their common environment, all expressed in terms of the reduced density
matrix, it is not such a clear-cut result. At least we have not seen a proof of it.

With the aim of assessing the validity of the CoM axiom the authors of \cite{CHY}  considered a system modeled by
$N$ harmonic oscillators  interacting with an environment consisting of $n$ harmonic oscillators  and derived an
exact non-Markovian master equation for such a system in a bath with arbitrary spectral density and temperature. The
authors outlined a procedure to find a canonical transformation to transform from the individual coordinates
$(x_i, p_i )$ to the collective coordinates $(\tilde{X}_i, \tilde{P}_i ), i=1,...,N$ where $\tilde{X}_1,
\tilde{P}_1$ are the center-of-mass coordinate and momentum respectively. In fact they considered a more general
type of coupling between the system and the environment in the form $f(x_i)q_j$ (instead of the ordinarily
assumed $x_i q_j$) and examined if the CoM coordinate dynamics separates from the reduced variable dynamics. They
noted that if the function $f(x)$ has the property $\sum_{i=1}^{N} f(x_i) = \tilde{f}(\tilde{X}_1) +
g(\tilde{X}_2,...,\tilde{X}_N)$, for example $f(x) = x$ or $f(x) = x^2$, one can split the coupling between the
system and environment into couplings containing the CoM coordinate and the relative coordinates. Tracing out the
environmental degrees of freedom $q_i$, one can easily get the influence action which characterizes the effect of
the environment on the system.

However, as the authors of \cite{CHY} emphasized, the coarse-graining made by tracing out the environmental
variables $q_i$ does not necessarily lead to the separation of the CoM and the relative variables in the
effective action. When they are mixed up and can no longer be written as the sum of these  two contributions, the
form of the master equation will be radically altered as it would contain both the relative variable and the
center-of-mass variable dynamics.

With these findings they conclude that for the N harmonic oscillators
quantum Brownian motion model, the coupling
between the system and the environment need be bi-linear, in the form $x_i q_j$,  for this axiom to hold. They
also proved that the potential $V_{ij}(x_i - x_j)$ is independent of the center-of-mass coordinate.  In that
case, one can say that the quantum evolution of a macroscopic object in a general environment is completely
described by the dynamics of the center-of-mass canonical variables ($\tilde{X}_1,\tilde{P}_1$) obeying a master
equation of the HPZ \cite{HPZ} type.

What is the relevance of this finding to MQP?  Within the limitations of the N harmonic oscillator model it
conveys at least two points: 1) For certain types of coupling the center of mass (CoM) variable of an object
composed of a large number of constituents does play a role in capturing the collective behavior of this object
2) Otherwise, more generally, the environment-induced quantum statistical properties of the system such as
decoherence and entanglement could be more complicated. (For a similar conclusion considering the cross level (of
structure) coarse-graining, see \cite{EntMm}.)

We next investigate the quantum entanglement between two macroscopic objects, comparing the entanglement between the
micro-variables of their constituents in two types of couplings: one-to-one and one-to-many. The very different
natures between these two types serve to illustrate the relevance of how the micro-constituents organize into a
macro object and how  entanglement between collective variables reveals the quantum features of a macroscopic
entity.

\subsection{Two different interaction patterns}
\label{sec:inter-intra}

We first apply the methods developed in \cite{CHY} to the study of the entanglement between the CoM coordinates of
two macroscopic objects. Each macroscopic object is modeled by $N$ identical coupled oscillators (NOS). However,
unlike \cite{CHY}, we do not include an environment in our discussion because our focus is on the entanglement
between the two objects induced by various types of direct interactions between their microscopic constituents.
We denote the coordinates and the momenta of the microscopic constituents of the two macroscopic
objects A and B by $\left\{ \xo{i},\po{i} \right\}$ and $\left\{ \xt{i} , \ptw{i} \right\}$ respectively. The interactions
between the microscopic constituents of one macroscopic object are assumed to be functions of the difference of
variables only and we restrict ourselves to bilinear couplings between the microscopic constituents of the two
macroscopic objects. The total Hamiltonian is then given by:

\begin{eqnarray}
  H_A &=& \sum_{i=1}^{N} \left( \frac{\po{i}^2}{2 M}
  +\frac{1}{2} M \om^2 \xo{i}^2 \right) + \sum_{i\ne j}^N
  \Vo{i}{j} ,\\
  H_B &=& \sum_{i=1}^{N} \left( \frac{\ptw{i}^2}{2 M}
  +\frac{1}{2} M \om^2 \xt{i}^2 \right) + \sum_{i\ne j}^N
  \Vt{i}{j} ,\\
  H_I &=& \sum_{i, j}^N \fg_{ij} \xo{i} \xt{j} .
\end{eqnarray}
The canonical transformation described in the Appendix A of \cite{CHY} can be applied to each
object separately to yield a new set of phase space variables $\left\{
\tXo{i}, \tPo{i} \right\}$ and $\left\{
\tXt{i}, \tPt{i} \right\}$ and the associated masses $\Mo{i}$. Here
$\tXo{1} = \frac{1}{N}\sum_{n=1}^{N} \xo{i}$ and
$\tXt{1}= \frac{1}{N}\sum_{n=1}^{N} \xt{i}$ are the CoM coordinates. The Hamiltonians of the macroscopic objects can
be written in terms of these variables as:

\begin{eqnarray}
  \label{H1v2}
  H_A &= \sum_{i=1}^{N} \left( \frac{\tPo{i}^2}{2 \Mo{i}}
  +\frac{1}{2} \Mo{i} \om^2 \tXo{i}^2 \right) + \tilde{V} \left(
  \tXo{2}, \cdots , \tXo{N}
  \right) = H_{A,CoM} + H_{A,REL} ,\\
  \label{H2v2}
  H_B &= \sum_{i=1}^{N} \left( \frac{\tPt{i}^2}{2 \Mt{i}}
  +\frac{1}{2} \Mt{i} \om^2 \tXt{i}^2 \right) + \tilde{V} \left(
  \tXt{2}, \cdots , \tXt{N}
  \right) = H_{B,CoM} + H_{B,REL} .
\end{eqnarray}

It has been proven in \cite{CHY} that the potential $\tilde{V}$ is not a function of the CoM
variable. This is a consequence of the form assumed for the potential energy. For a general bilinear coupling
characterized by $\fg_{ij}$ the interaction Hamiltonian $H_I$ can take on a complicated form, possibly mixing the CoM
variables with the relative variables.

In what follows we will focus on two particular choices of $\fg_{ij}$, inspired by Martins \cite{Martins}. The use of the new set of canonical variables which include the CoM will help interpret the behaviour of macroscopic entanglement.

\subsubsection{Pairwise interaction pattern}
\label{sec:pairwise}

The pairwise interaction pattern is defined by $\fg_{ij} = \og
\delta_{ij}$ (see Fig. 1(a)). In other words one constituent particle modeled by an oscillator from object A couples to one oscillator from object B, assuming that all pairwise couplings have the same strength. Using the canonical transformation of \cite{CHY} it can be shown that the interaction Hamiltonian takes the form:
\begin{eqnarray}
  \label{pairwise}
  H_I &= \sum_{i}^{N} \frac{\og}{M} \Mo{i} \tXo{i} \tXt{i} .
\end{eqnarray} Note that pairwise interactions among the original variables translate into pairwise interactions
among the transformed variables. A very important difference is that whereas the pairwise interactions in the
original variables were all of equal strength, the strength of the interactions scale with the mass of the variables
after the transformation. As a result the relative strength of interactions between variable pairs are the same
for all the variables, including the CoM. To see this explicitly let us consider the case with $\tilde{V}=0$ for
simplicity, namely the micro-constituents of each macroscopic object do not interact with each other. Then we
rescale the coordinates by $\tXo{i}\rightarrow \bXo{i}
\sqrt{M/\Mo{i}}$ and $\tXt{i}\rightarrow \bXt{i}
\sqrt{M/\Mt{i}}$, after which the Hamiltonian takes the form:
\begin{eqnarray}\label{rescaled}
  H &= \sum_{i=1}^{N} \left( \frac{\bPo{i}^2}{2 M} + \frac{1}{2}M \om^2
  \bXo{i}^2 +
  \frac{\bPt{i}^2}{2 M} + \frac{1}{2}M \om^2 \bXt{i}^2 + \og \bXo{i} \bXt{i}
  \right) .
\end{eqnarray} In this form it is easy to see that the effective
strength of interactions in the CoM variable is
the same as the effective strength of interactions in all the other variables. Hence the pairwise interaction
pattern will induce the same amount of entanglement between pairs of transformed variables, without
distinguishing the CoM variable. Entanglement between non-CoM variables would be effected if the interactions
among the oscillators within the same object, i.e. $V_{ij}$, are not set to zero.

\begin{figure}
\begin{center}
\includegraphics[scale=0.45]{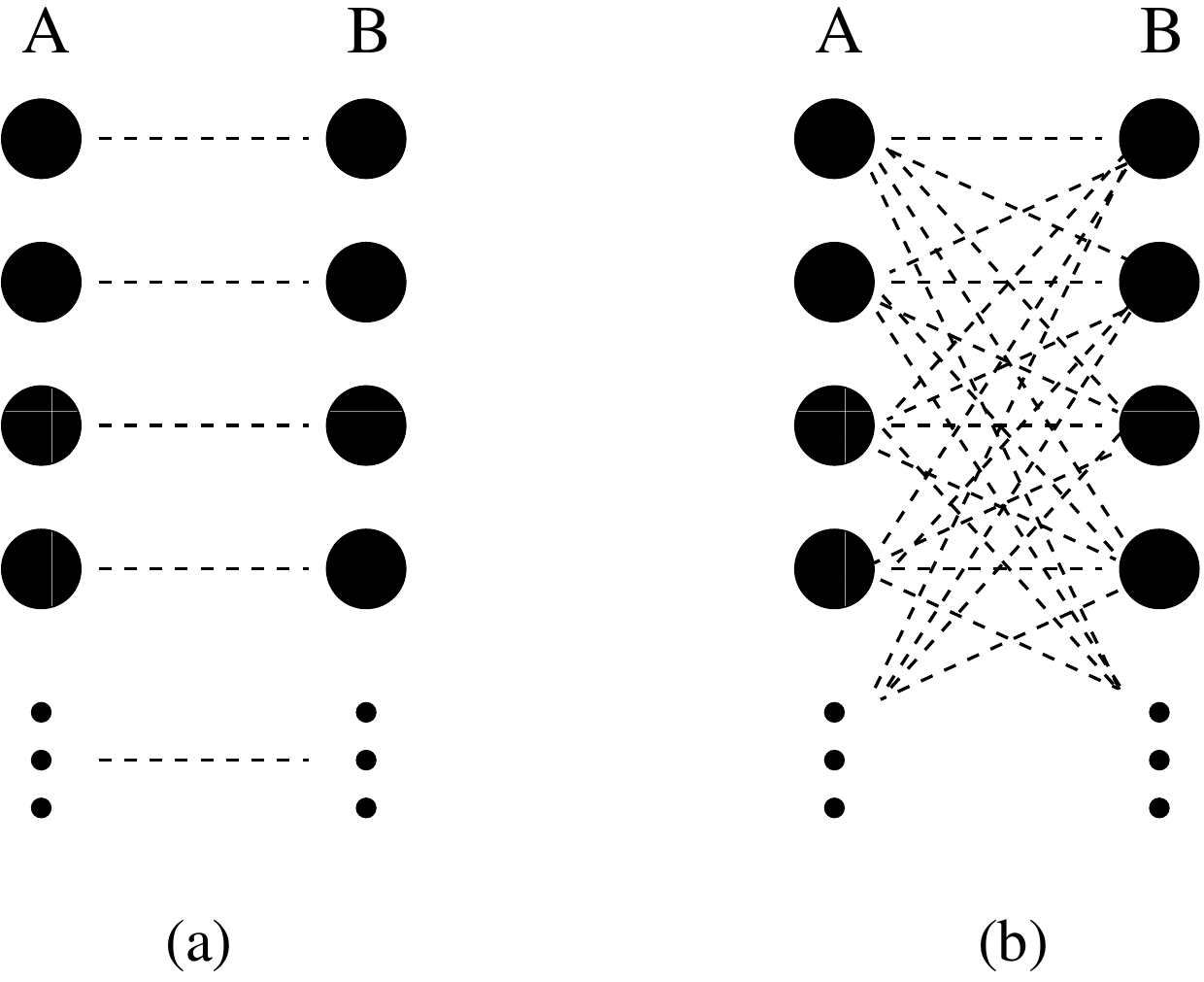}
\label{fig:pattern}
\caption{Schematic representation of the two types of couplings
studied in this paper: (a) Pairwise interaction pattern (b) One-to-all
interaction pattern}
\end{center}
\end{figure}
If we only focus on the effect of the pairwise interactions, it is fair to say that such interactions couple the
pairwise transformed variables with equal effective strength independent of the size $N$ of the macroscopic
objects. As a consequence we expect the behavior of entanglement between the corresponding variables of the
objects to be independent of the size of the macroscopic objects, even for the CoM coordinate. For
instance, at a given temperature the amount of entanglement between the two corresponding variables of the
objects will not depend on $N$. The critical  temperature above which the entanglement ceases to exist also
should not depend on $N$.

\subsubsection{One-to-all interaction pattern}
\label{sec:onetoall}

The one-to-all interaction pattern is characterized by $\fg_{ij}= \og$
(see Fig. 1(b)). Then it is easy to see that the interaction
Hamiltonian in the transformed variables takes the form: \begin{eqnarray}
  \label{onetoall}
  H_I &= N^2 \og \tXo{1} \tXt{1} .
\end{eqnarray} Note that \ota interaction pattern corresponds to a coupling only between the CoM variables of the macroscopic objects, the relative variable Hamiltonian is unaffected. Thus \ota pattern differs from the pairwise
pattern in that it distinguishes the CoM variable. Moreover if we perform the same rescaling of the previous
section to determine the effective strength of this coupling we get:
\begin{eqnarray}
  \label{rescaled2}
  H_{CoM} = \frac{\bPo{1}^2}{2 M} + \frac{1}{2}M \om^2 \bXo{1}^2 +
  \frac{\bPt{1}^2}{2 M} + \frac{1}{2}M \om^2 \bXt{1}^2 + N \og \bXo{1}
  \bXt{1} .
\end{eqnarray} We see that the effective strength of the coupling increases with increasing $N$ for the \ota
pattern. Thus in this case we expect the entanglement between the CoM variables to increase with increasing size
of the macroscopic objects and survive at higher temperatures. The \ota interaction pattern is crucial for the
scaling of the entanglement of CoM variables with $N$. Hence it is important to investigate if this type of
interaction pattern can occur in realistic situations and if so how
common it is. In the Discussion section we will present a different
perspective in understanding the one-to-all coupling case, as
characterizing different identical constituents at the same level of
structure of one macro-object rather than between the two objects.

\section{Conditions for CoM variable to decouple and its role in MQP}

In this section we derive the necessary and sufficient conditions for
the CoM variables of two macroscopic objects to decouple from the
rest of the degrees of freedom. The macroscopic objects are modeled by N coupled oscillators
and interact via bilinear couplings.
Eqs.(\ref{H1v2},~\ref{H2v2}) show that $H_A$ and $H_B$ obey the CoM
axiom independently. Below, we demonstrate under which conditions $H_I = H_{I,CoM}
+ H_{I,REL}$ where $H_{I,CoM}$ is a function of the CoM coordinates of
both macroscopic objects only and $H_{I,REL}$ does not depend on the
CoM coordinates of either object. Specifically, we will derive the conditions under which
\begin{eqnarray}
  \nonumber
  H_I &=& H_{I,CoM}(\tXo{1},\tXt{1}) + H_{I,REL}(\tXo{2},\cdots,
\tXo{N}, \tXt{2},\cdots, \tXt{N})\\
&=& \sg_{11} \tXo{1} \tXt{1} + \sum_{i,j\ne 1} \sg_{ij}
\tXo{i} \tXt{j}
\label{H_Iform}
\end{eqnarray}
To this end we follow the strategy adopted in Appendix C of
Ref.\cite{CHY} and determine the functional form of $H_I$ by
calculating its partial derivatives.
\begin{eqnarray}
  \frac{\partial H_I}{\partial \tXo{1}} = \sum_{i,j} \fg_{ij}
  \left(\frac{\partial \xo{i}}{\tXo{1}}\right) \xt{j} =
 \sum_j \xt{j} \left( \sum_i \fg_{ij} \right)
\end{eqnarray}
where in the second equality we used $\partial \xo{i}/\partial\tXo{1} =
1$ for all $i$, which can be shown by explicitly constructing a
coordinate transformation whereby one coordinate is the CoM coordinate.
For an example consider the construction described in the Appendix B of \cite{CHY} or the explicit construction described in Section
\ref{sec:A} of this paper, albeit with a different normalization for the CoM
coordinate.
Since we want $H_I$ to have the form given by Eq.(\ref{H_Iform}) we
require $N \sum_i \fg_{ij} = \sg_{11}$, which is independent of $j$.
Repeating this derivation by replacing subscript $A$ with $B$ we
obtain the second condition that $N \sum_j \fg_{ij}= \sg_{11}$, which is independent of $i$.
To summarize, the necessary and sufficient conditions for the CoM
variables of both macroscopic objects to decouple from the relative coordinates is:
\begin{eqnarray}
   \sg_{11} = N \sum_i \fg_{ij} = N \sum_j \fg_{ij}.
\end{eqnarray}
If we do the same rescaling as in Eq.~\eqref{rescaled} we obtain:
\begin{eqnarray}
  H_{I,CoM} &=& \bar{G}_{11} \bar{X}_{A1} \bar{X}_{B2}\\
  \label{effstrength}
  \bar{G}_{11} &=& \sum_{i=1}^{N} g_{ij} = \sum_{j=1}^{N} g_{ij}
\end{eqnarray}

As a quick check it can be easily verified that both the ``pairwise'' and ``one-to-all'' couplings satisfy this condition with
$\bar{G}_{11}= \og$ and $\bar{G}_{11}=N \og$ respectively.
These results agree with Eq.~\eqref{rescaled} and Eq.~\eqref{rescaled2}.

Up to this point we have only discussed two different patterns of couplings in
detail, the pairwise and one-to-all.
By analyzing Eq.~\eqref{effstrength} we can see how the effective
coupling strength of the CoM variables will behave for different
patterns. For instance, if the summations converge as $N\rightarrow
\infty$ we conclude that in the thermodynamic limit the effective
coupling strength of the CoM variables is an intrinsic quantity,
independent of the size of the system. The only interaction pattern
for which the effective coupling strength is extensive is the
one-to-all pattern. Any other pattern for which the summation in
Eq.~\eqref{effstrength} is divergent corresponds to an effective
coupling that increases with the system size. If $|g_{ij}| >
|g_{ik}|$ for $k>j>i>$, this corresponds to a sub-linear growth. For
example:
\begin{eqnarray}
  g_{ij} &=& \frac{\lambda}{(i-j)_{mod\, N}+c_1}, \quad \mathrm{where}\, c_1 \,
  \mathrm{is\, a\,
  constant}, \\ 
  \bar{G}_{11} &=& \lambda \sum_{i=1}^{N} \frac{1}{(i-j)_{mod N}+c_1} = \lambda
  \sum_{i=0}^{N-1} \frac{1}{i+c_1}  \xrightarrow[\lim N
  \rightarrow \infty]{} \log(N) + c_2,
\end{eqnarray}
where $c_2$ is a constant that depends on $c_1$ and we used the
convention $(-i)_{mod N} = N-i$ for $i<N$. The reason we used
$(i-j)_{mod\, N}$ in the interaction term is in order to satisfy the
condition Eq.~\eqref{eq:cond3}.
We see that in the
thermodynamical limit the effective interaction strength scales as
$\log(N)$.

\subsection{Explicit Canonical Transformation for $N=2^\n$}
\label{sec:A}

\begin{figure}
  \begin{center}
  \includegraphics[scale=0.24]{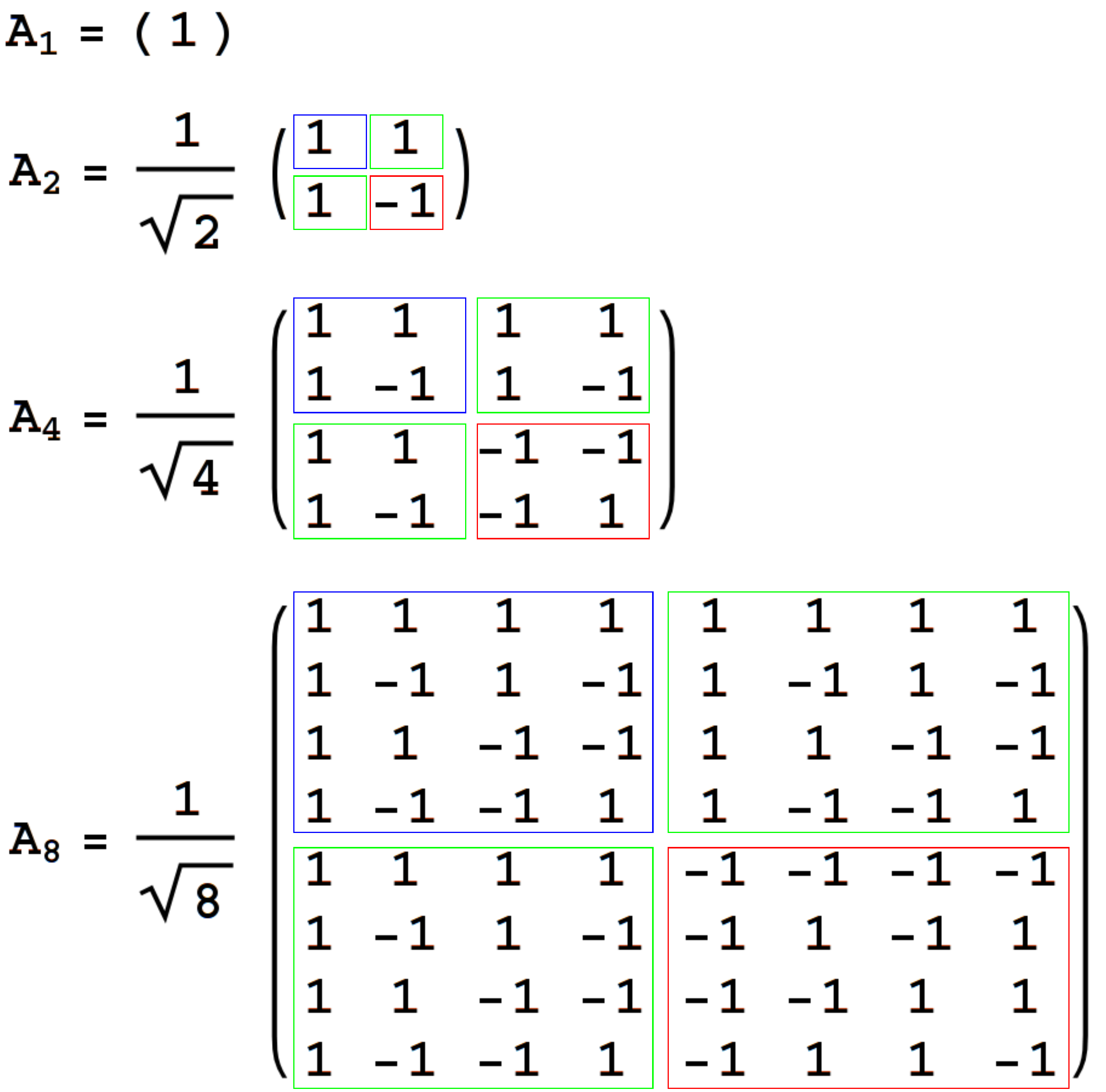}
\label{A}
\caption{$\mathbf{A}_N$ is the transformation matrix from the original
coordinates to the new set of coordinates which include the CoM. A procedure to explicitly
construct $\mathbf{A}_N$ is illustrated above for $N=2^\n$. If
$\mathbf{A}_{2^\n}$ is known, $\mathbf{A}_{2^{\n+1}}$ can be
constructed by copying the block matrix $\mathbf{A}_{2^\n}$ to the
the off-diagonal blocks and by putting the negative of it to the lower
diagonal. Note that any transformation matrix for a smaller $N$ can be
obtained by restricting to the upper left corner of the larger matrix.}
\end{center}
\end{figure}

In this section we describe a change of coordinates from the original
set $\left\{ \xo{i},\xt{i} \right\}$ into a new set $\left\{
\Xo{i},\Xt{i}
\right\}$. This new canonical transformation is more symmetric than the one used in
the previous section and in \cite{CHY}, and allows for a general
analysis for $N=2^\n$, where $\n$ is an arbitrary integer.
Since we are mainly interested in the behavior of the CoM coordinate
(and how it differs from the rest of the coordinates),
we require that the new set includes two coordinates
$\Xo{1}$ and $\Xt{1}$, which correspond to the CoM of
objects $A$ and $B$.
\footnote{Note however that these are not the standard CoM coordinates but are
rescaled by a factor of $\sqrt{N}$, i.e.
$\Xo{1}=(\xo{1}+\cdots+\xo{N})/\sqrt{N}$. See Figure~\ref{A} for the
definition of the rest of the coordinates.
This rescaling is purely conventional and does not effect the
physical conclusions drawn about the CoM.}

This set of new coordinates allow us to generalize our previous analysis to include
randomness in the couplings between the microscopic constituents of
objects $A$ and $B$ which is shown in the next subsection.
Our results  show that the CoM variable is singled out by i.i.d.
(independent and identically distributed) random couplings between the
micro-variables. Moreover, the properties of the canonical transformation provide insights into
the reason why the CoM variable is special.

To distinguish this set of coordinates from those defined in Section \ref{sec:inter-intra} we
drop the tilde. The transformation described below is only valid for $N=2^\n$.
However, this is enough for our purpose of addressing MQP for large $N$.

Let $\mathbf{A}_N$ be the matrix associated with the linear
transformation from the original coordinates $\mathbf{x}=\left\{ x_i \right\}$ to
the new coordinates $\bfX=\left\{ \nX_i
\right\}$. In the rest of the paper we denote vectors and matrices
with bold characters, whereas individual entries will be indicated by
regular characters with subscripts. Then
\begin{equation}
  \label{trans}
\bfX=\mathbf{A}_N \cdot \mathbf{x}.
\end{equation}
Explicit form of $\mathbf{A}_N$ for $N=1,2,4,8$ is given in Figure
\ref{A} and the procedure for obtaining $N=2^\n$ for arbitrary integer
$n$ is described. For brevity of notation the subscript $N$ will
be dropped in the rest of this section. A nice property of the transformation matrix is
that $\mathbf{A}^{-1}=\mathbf{A}=\mathbf{A}^T$.
Note that $\mathbf{A}$ has the first column (row) of identical
entries, which corresponds to the CoM coordinate. The relative coordinates
defined in this section are different from those used in Section
\ref{sec:inter-intra} as defined in \cite{CHY}.
The coordinates of this section are in a way more symmetric; for
instance all the associated masses are equal to the original mass $M$.
Demanding this symmetry together with the condition that there are two coordinates proportional to the CoM of each
object forces a specific form on the matrix $\mathbf{A}$:
all the columns (rows) corresponding to the relative coordinates have half of the entries with positive
and the other half negative sign.
This property, which singles out the CoM coordinate kinematically, will play an important
role in the proceeding analysis.

Using the transformation \eqref{trans} we can write the Hamiltonian
for the new set of canonical variables. In this section we will set
$V_{ij}=0$ since we are interested in the effect of interactions
between the constituents of the two macroscopic objects. The
Hamiltonians $H_{A}$ and $H_B$ preserve their original form under the
transformation since $\mathbf{A}^T\mathbf{A}=\mathbf{1}$. The
interaction Hamiltonian becomes:
\begin{equation}
  H_I = \sum_{ij} \Xo{i} \tg_{ij} \Xt{j}, \qquad \tgm =
  \mathbf{A} \cdot \fgm \cdot \mathbf{A}.
  \label{eq:cond3}
\end{equation}
As a quick check of this formalism we calculate $\tgm$ for
the pairwise and one-to-all interaction patterns studied in Sections
\ref{sec:pairwise} and \ref{sec:onetoall}. For pairwise interactions
$g_{ab} = \og \delta_{ab}$ and we
get $\tg_{ij} = \og \sum_{ab} A_{ia} \delta_{ab}
A_{bj}= \og
\left(\mathbf{A}^2\right)_{ij}= \og \mathbf{1}_{ij}= \og \delta_{ij}$. Note that each coordinate is coupled with equal strength in this type of
coupling (there was no need to renormalise the coordinates since the
associated masses are already equal in this set of coordinates).
For one-to-all
interactions we have $\fg_{ab} = \og$ which translates to $\tg_{ij} = \og
\sum_{ab} A_{ia} A_{bj} = \og \left(\sum_a
A_{ia}\right) \left( \sum_b A_{bj}\right) = N \og
\delta_{i1}\delta_{1j}$.
Note that only the CoM coordinates are coupled in this type of
interaction and the coupling strength scales as $N$, which agrees with
previous analysis.

\subsection{Independent and Identically Distributed Gaussian Couplings}

In previous sections we treated coupling patterns that are
deterministic. It is reasonable to ask whether the conclusions we
reach about the significance of the CoM coordinate and its decoupling
from the relative coordinates are stable under perturbations. To
investigate this issue, we reconsider the one-to-one and \ota patterns and
this time allow for Gaussian variations around the non-zero coupling
strengths. Our analysis shows that the conclusions of previous sections regarding the significance of the CoM variable and the decoupling of it from the relative coordinates are not altered by the addition of fluctuations.

Note that in what follows we do not allow for fluctuations in the
vanishing coupling strengths, for example the non-pairwise coupling
strengths in the one-to-one pattern (see Eq.~\eqref{eq:variancee}). We
motivate this choice by noting that the vanishing couplings can be the
result of a constraint based on symmetry or geometry and thus immune
to fluctuations. On the other hand, to assume that the values of non-zero
coupling constants are fixed without fluctuations would be more
difficult to justify.

\subsubsection{One-to-all Pattern}

With the canonical transformation of the previous section we can address the case where the coupling constants
$g_{ij}$ are sampled from identical independent Gaussian distributions
characterised by the mean and variances:
\begin{eqnarray}
  \langle \fg_{ab} \rangle &=& \bar{\fg}\\
  \label{eq:variance_all}
  \langle
\fg_{ab} \fg_{cd} \rangle -\langle \fg_{ab}\rangle \langle
\fg_{cd}\rangle &=& \delta_{ac} \delta_{bd} \sigma_
\fg^2\
\end{eqnarray}
We now ask the question, how do the coupling constants
$G_{ij}$ behave?
We can use the transformation \eqref{trans} and the properties of the matrix
$\mathbf{A}$ to calculate the statistical properties as:
\begin{eqnarray}
  \langle \tg_{ij} \rangle &=& \sum_{ab} \nA_{ia} \langle
  \fg_{ab} \rangle \nA_{bj} \nonumber \\
  &=&\bar{\fg} \left( \sum_a \nA_{ia}
  \right)\left( \sum_{b} \nA_{bj} \right) = N \bar{\fg} \delta_{i1}
  \delta_{j1}\\
  \langle \tg_{ij} \tg_{kl} \rangle &=& \langle
  \sum_{ab} \nA_{ia} \fg_{ab} \nA_{bj} \sum_{cd}
  \nA_{kc} \fg_{cd} \nA_{dl} \rangle \nonumber \\
  &=& \sum_{abcd}
  \nA_{ia}\nA_{bj}\nA_{kc}\nA_{dl} \left(
  \bar{\fg}^2+\delta_{ac} \delta_{bd} \sigma_\fg^2 \right)\nonumber \\
  &=& N^2 \bar{\fg}^2 \delta_{i1}\delta_{j1} \delta_{k1} \delta_{l1}
  +\sigma_\fg^2 \left(\mathbf{A}^2\right)_{ik}
  \left(\mathbf{A}^2\right)_{jl}\nonumber \\
  \langle \tg_{ij} \tg_{kl} \rangle - \langle
  \tg_{ij} \rangle \langle \tg_{kl} \rangle &=&
  \sigma_\fg^2 \delta_{ik} \delta_{jl} \label{eq:derivation1}
\end{eqnarray}
Thus if the couplings between the constituents of the macroscopic
objects are statistically independent Gaussian random variables, the
corresponding couplings between the new variables are also independent
Gaussian random variables, which follows from the fact that the new
and old variables are related by a linear transformation.
The main difference is that only the CoM-to-CoM coupling has a nonvanishing
expectation value which is equal to the expectation value of the
couplings of the original coordinates multiplied by $N$. As
expected, the average behavior is that of the deterministic rule
around which we are perturbing.
On the other hand the coupling constants of all of the
new coordinates, i.e. both CoM and relative coordinates, have the same
variance that is the same as the variance of the couplings of the
original coordinates.
Note that the fluctuations of the CoM coupling become negligible in
the thermodynamic limit, but not those of the relative
coordinates. Thus the conclusions of previous sections about the CoM hold with respect to
the perturbations considered here and in the thermodynamic limit.

\subsubsection{One-to-one Pattern}

Here we repeat the analysis of the previous section for the one-to-one
coupling pattern.
The coupling constants
$g_{ij}$ are sampled from identical independent Gaussian distributions
characterised by the mean and variances:\footnote{Note the difference
with Eq.~\eqref{eq:variance_all} in the variance term. As mentioned before, we allow for fluctuations of
couplings with non-vanishing means only.}
\begin{eqnarray}
  \langle \fg_{ab} \rangle &=& \bar{\fg} \delta_{ab}\\
  \langle
\fg_{ab} \fg_{cd} \rangle -\langle \fg_{ab}\rangle \langle
\fg_{cd}\rangle &=& \delta_{ab} \delta_{ac} \delta_{bd} \sigma_
\fg^2\ \label{eq:variancee}
\end{eqnarray}
We now ask the question, how do the coupling constants
$G_{ij}$ behave? After some algebra
we get:
\begin{eqnarray}
   \langle \tg_{ij} \rangle &=& \sum_{ab} \nA_{ia} \langle
  \fg_{ab} \rangle \nA_{bj} = \bar{\fg} \left( \sum_a \nA_{ia} \nA_{aj} \right) = \bar{\fg}
  \delta_{ij} \label{eq:mean2}\\
  \langle \tg_{ij} \tg_{kl} \rangle &=& \langle
  \sum_{ab} \nA_{ia} \fg_{ab} \nA_{bj} \sum_{cd}
  \nA_{kc} \fg_{cd} \nA_{dl} \rangle \nonumber \\
  &=& \sum_{abcd}
  \nA_{ia}\nA_{bj}\nA_{kc}\nA_{dl} \left(
  \bar{\fg}^2 \delta_{ab} \delta_{cd}+\delta_{ab} \delta_{ac}
  \delta_{bd} \sigma_\fg^2 \right)\nonumber \\
  &=& \bar{\fg}^2 \delta_{ij}\delta_{kl}
  +\sigma_\fg^2 \sum_a
  \left(\mathbf{A}_{ia}\mathbf{A}_{ja}\mathbf{A}_{ka}\mathbf{A}_{la}\right)
  \nonumber \\
  \left| \langle \tg_{ij} \tg_{kl} \rangle - \langle
  \tg_{ij} \rangle \langle \tg_{kl} \rangle \right| &\le&
  \sigma_\fg^2 / N. \label{eq:variance2}
\end{eqnarray}
In the last step we used the fact that the matrix elements of
$\mathbf{A}$ are $\pm
1/\sqrt{N}$ to conclude that $\left| \sum_a
\left(\mathbf{A}_{ia}\mathbf{A}_{ja}\mathbf{A}_{ka}\mathbf{A}_{la}\right)\right|
\le 1/N$.
We see that, unlike the one-to-all case, the CoM
coupling behaves the same as those of the relative coordinates and for
CoM as well as relative coordinates the fluctuations become negligible in the thermodynamic limit.

\section{Inter- and Intra- Level-of-Structure Entanglement in 4 Coupled-Oscillator System}

In the above section we showed how the scaling with N depends on the type of couplings amongst the constituent particles of two objects and provided formal proofs for the decoupling of the center of mass variables.  In this section we use an explicit model calculation to illustrate the ideas presented in prior sections, namely, the level of structure scheme in the consideration of entanglement of composite objects, and the special role of the CoM variables in the consideration of macroscopic entanglement.  The model is made up of 4 coupled oscillators in two pairs representing the two objects A (constituents 1, 2) and B (constituents 3, 4) \footnote{In the level of structure scheme (A, B) can be viewed as at a higher level of structure (more composite in nature, such as nucleons) than (1,2) or (3, 4) at a lower level of structure (more 'elementary', such as quarks). To contrast the two levels of structure we may refer to them as  `macro' vs micro-objects with no harm done.}. We assign different coupling strengths: $\alpha$ as that between the two (intra-level) constituents 1-2 or 3-4  in each object, $\beta$ as the (inter-level) coupling strength between 1-3 and also 2-4 (parallel), and $\gamma$ between 1-4 and also 2-3 (cross).  Considering $\alpha$ to be large compared to $\beta$ or $\gamma$ refers to strong intra- level interactions or tight-binding of constituents within one level of structure. This model can show the difference between the cases studied in Sec. 3, namely, the pairwise (1-3, 2-4) interaction versus the one to all (1-3 and 1-4) interaction.  As in Sec. 3 we see that it is only in the one-to-all case that the entanglement scales with N. After this we calculate the entanglement dynamics of the 4-oscillator system with disparate coupling strength, while distinguishing three cases of entanglement: The case between the center of mass variables (CoM) of A and B,  the case between the CoM variable of one object A and one constituent particle of B, and the case between one constituent particle of A and one of B. We see some pariwise entanglement meets sudden death while others sustain over long time ranges. This offers another way to determine under what conditions quantum entanglement between macroscopic objects can persist.

\subsection{The 4 oscillator model}

The total Hamiltonian of the system of 4 coupled harmonic oscillators of equal masses but disparate coupling strength as described above is given by
\begin{eqnarray}
  H &=&\sum_{a=\left\{ A,B \right\}}\sum_{j=\{ 1,2\}}
  (\frac{1}{2}\dot{x}_{a j}^2 +
  \frac{k}{2}x_{a j}^2) + \alpha(x_{A 1}x_{A 2} + x_{B 1}x_{B 2})
  \\
  &&\hspace{22mm}  + \beta(x_{A 1}x_{B 1}x_{A 2}x_{ B 2}) \nonumber  +\,  \gamma(x_{A 1}x_{ B 2} + x_{A 2}x_{B 1}) \nonumber \\
 &\equiv& T + U
\end{eqnarray}
where $T=\sum_{a,j}
\frac{1}{2}\dot{x}_{a j}^2,
U=\frac{1}{2}\sum_{a, j , a',j'} x_{a j} V_{a j
a' j'} x_{a j'}$ and we have set the mass of each oscillator to unity. We use indices $a, a'$ to label the objects A, B and $j, j'$ to label the constituent particles. The  $u_{ij}$
characterizes the interactions among the oscillators and has the
matrix form
\begin{eqnarray}
  \mathbf{V} = \left(
              \begin{array}{cccc}
                k & \alpha & \beta & \gamma \\
                \alpha & k & \gamma & \beta \\
                \beta & \gamma & k & \alpha \\
                \gamma & \beta & \alpha & k \\
              \end{array}
            \right)
\end{eqnarray}
This Hamiltonian can be diagonalized into the form
\begin{eqnarray}
H &=&\sum_{i=1}^4 (\frac{1}{2}\dot{\eta}_i^2 +
\frac{\omega_i^2}{2}\eta_i^2),
\end{eqnarray}
where $\omega_1^2 = k +
\alpha+\beta+\gamma,\omega_2^2=k+\alpha-\beta-\gamma,\omega_3^2=k-\alpha+\beta-\gamma,\omega_4^2=k-\alpha-\beta+\gamma$.
The $\eta_i$ could be written in terms of the center-of-mass
($X_{A 1}=\frac{1}{2}(x_{A 1}+x_{A 2})$) and relative coordinates
($X_{A 2}=x_{A 1}-x_{A 2}$) of object A and similarly for object B. The explicit
form of $\eta_i$ are
\begin{eqnarray}
  \eta_1 &=&
  \frac{1}{2}(x_{A 1}+x_{A 2}+x_{B 1}+x_{B 2})=X_{A 1}+X_{B 1}  \nonumber \\
\eta_2 &=& \frac{1}{2}(x_{A 1}+x_{A 2}-x_{B 1}-x_{B 2})=X_{A 1}-X_{B 1}  \nonumber \\
\eta_3 &=& \frac{1}{2}(x_{A 1}-x_{A 2}+x_{B 1}-x_{B
2})=\frac{1}{2}(X_{A 2}+X_{B 2})  \nonumber \\
\eta_4 &=& \frac{1}{2}(x_{A 1}-x_{A 2}-x_{B 1}+x_{B
2})=\frac{1}{2}(X_{A 2}-X_{B 2})
\end{eqnarray}

As a simple illustration we assume that at time ($t=0$) the couplings among the four oscillators were zero and the initial
wave function of the system was the direct product of the Gaussian ground state
wave function of each oscillator, i.e.,
$\Psi(x_{A 1},x_{A 2},x_{B 1},x_{B 2};0)=\psi_0(x_{A 1})\psi_0(x_{A
2})\psi_0(x_{B 1})\psi_0(x_{B 2})$, where $\psi_0(x)= (\frac{\omega_0}{\pi \hbar})^{1/4}\exp(-\frac{\omega_0 x^2}{2\hbar})$. At time $t >0$ the interactions among the oscillators were turned on and the
total system is then driven by the Hamiltonian $H$ and the final
wave function at time $t$ is
$\Psi(x_{A 1},x_{A 2},x_{B 1},x_{B 2};t)$ which can be easily obtained from the evolution operator of the system.

\subsection{Entanglement measure for continuous variables}

There are well defined measures of entanglement for continuous variables. Define the two-point correlation matrix $V(Q_j,Q_i)$ with elements
\begin{equation}
  V_{\mu\nu}(Q_j,Q_i)(t)=\left<\right.{\cal R}_\mu ,{\cal R}_\nu \left.\right>
    \equiv {1\over 2}\left<\right. \left( {\cal R}_\mu {\cal R}_\nu +
    {\cal R}_\nu {\cal R}_\mu \right) \left.\right> \label{CorrMtx}
\end{equation}
where ${\cal R}_\mu = (\hat{Q}_j, \hat{P}_j, \hat{Q}_i, \hat{P}_i)$,
$\mu, \nu = 1,2,3,4$. Note that here $\hat{P}_i$ is the conjugate momentum operator of $\hat{Q}_i$ and the $Q$ could be
$x, \eta , X,$ or $R$.
The partial transpose of ${\bf V}(Q_j,Q_i,t)$ is ${\bf V}^{PT}(Q_j,Q_i,t) = {\bf \Lambda V}(Q_j,Q_i,t) {\bf
\Lambda}$, where ${\bf \Lambda} ={\rm diag}( 1,1,1,-1)$. Starting with
the Gaussian initial state, $\Psi(x_{A1},x_{A2},x_{B1},x_{B2},0)$ , the reduced density
matrix or Wigner function of the two oscillators is always Gaussian by virtue
of the quadratic structure of the Hamiltonian of our model. Therefore the oscillator $Q_i$ and
oscillator $Q_j$ is entangled at time $t$ if and only if \cite{PPTSimon}
\begin{equation}
  \Sigma(Q_j,Q_i,t) \equiv \det \left[ {\bf V}^{PT}(Q_j,Q_i,t)+
  i{\hbar\over 2}{\bf M} \right] < 0  \label{Entcond}
\end{equation}
where
\begin{equation}
  {\bf M} \equiv \left( \begin{array}{cccc}
    0 & 1 & 0 & 0 \\
    -1 & 0 & 0 & 0 \\
    0 & 0 & 0 & 1 \\
    0 & 0 & -1 & 0 \end{array} \right)
\end{equation}
is a symplectic matrix. Note that when $\Sigma \le 0$ the behavior of
$\Sigma$ is quite similar to the behavior of the negative eigenvalues
which are connected to the logarithm negativity \cite{VW02}. Thus the
value of $\Sigma$ itself is a good indicator of the degree of
entanglement, at least in this specific model. It is then straightforward (although somewhat tedious) to obtain the analytic form
of $\Sigma(Q_j,Q_i,t)$ and hence the entanglement dynamics.

Below we calculate the two-point correlation matrix $V$ and $\Sigma$ for various degrees of freedom in this model and discuss the implications
on MQP.

\subsection{Entanglement Dynamics with Pairwise Disparate Coupling}

We now study the entanglement structure and its dynamics for different pairs of variables, those which represent the objects A, B -- specifically the CoM variables $X_A, X_B$ and the relative coordinates $R_A, R_B$ -- and those representing the constituents particles $(1, 2), (3, 4)$. In this way we can see how these variables  and how the inter-level and intra-level coupling strengths enter into the entanglement structure. The results ranges from the somewhat obvious, such as that $\Sigma(X_A,R_A,t)=\Sigma(X_A,R_B,t), \Sigma(X_A,X_B,t)=2\Sigma(X_A,x_{B1},t),$ and $\Sigma(X_A,x_{B1},t)=\Sigma(R_A,x_{B1},t)$ to the more complicated and subtle, such as the existence of sudden death for certain pairs of variables and others which persists, as illustrated in the figures.

We divide into two categories, that when the intra-level coupling strength $\alpha$ is zero or nonzero. One can think of the former as depicting systems with very weak coupling amongst the constituent particles (molecular gas) and the latter as objects with stronger constituent couplings (solid body).  We will present the formal results in the following and attempt to understand their physical relevance in the next section.  Note that $\alpha =0$ is a special case for Sec. 3.

\subsubsection{Entanglement Dynamics for  $\alpha=0$}

In this category there is no direct interactions between the constituents of each object. We will then concentrate on the interactions between objects A and B \footnote{In the spirit of the level-of-structure scheme, we may also refer to them as at a higher level of structure, or simply subsystems A and B.} There are two special subcases:

\begin{itemize}

\item  $\beta \neq 0, \gamma=0$: pairwise interaction case. Oscillator 1(2) in subsystem A couples only to oscillator 3(4) in subsystem B.
Fig. 3(a) shows the entanglement dynamics of $\Sigma(X_A,x_{B1},t), \Sigma(x_{A1},x_{B1},t)$ and Fig. 3(b) shows the entanglement dynamics of $ \Sigma(X_A,X_B,t), \Sigma(X_A,R_B,t), \Sigma(R_A,R_B,t)$. Note that $\Sigma(X_A,R_B,t)=\Sigma(X_A,R_A,t) \geq 0$ in this case and hence there is no entanglement among $X_A$ and $R_B(R_A)$.

\begin{figure}
\includegraphics[width=6cm]{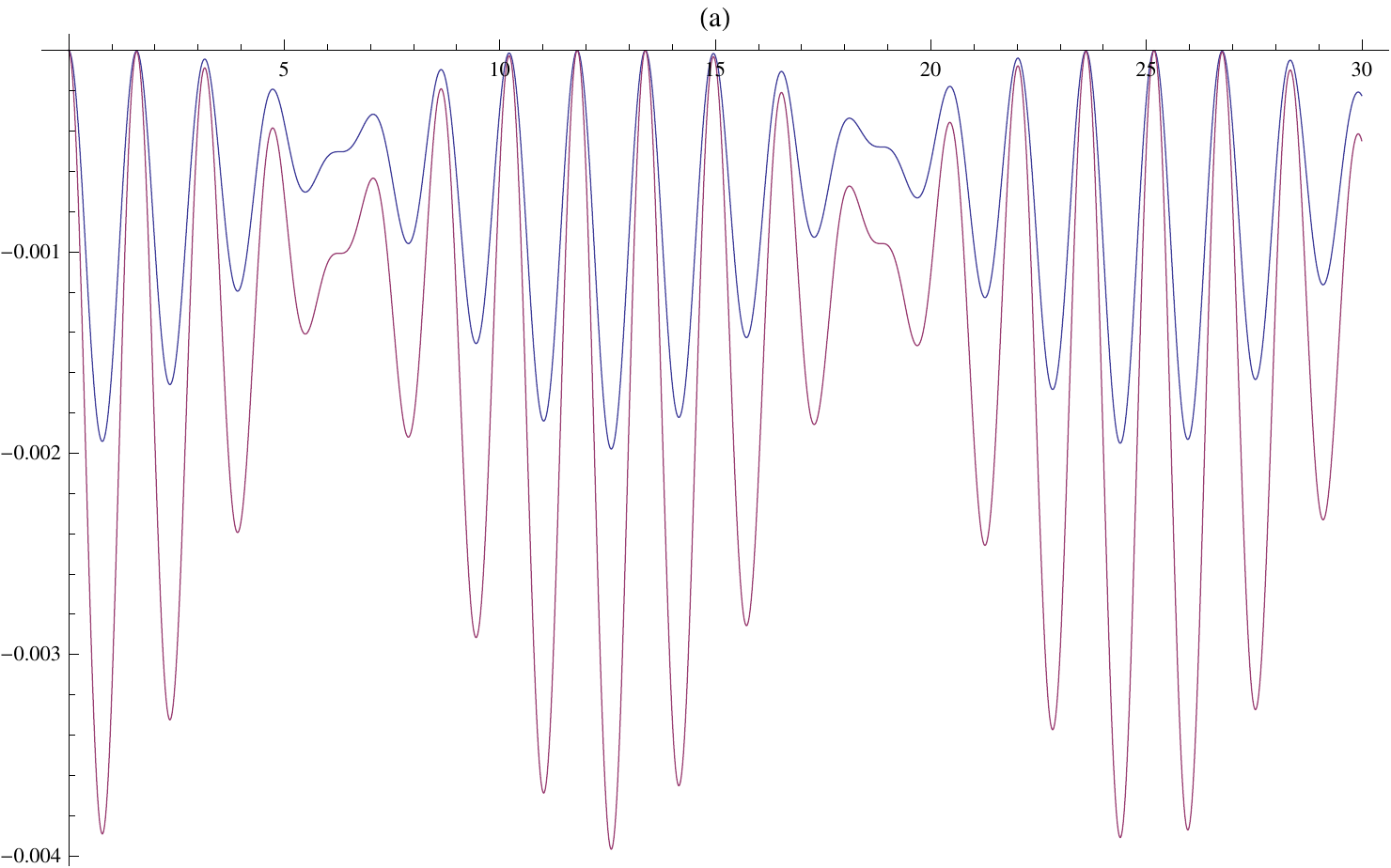}
\includegraphics[width=6cm]{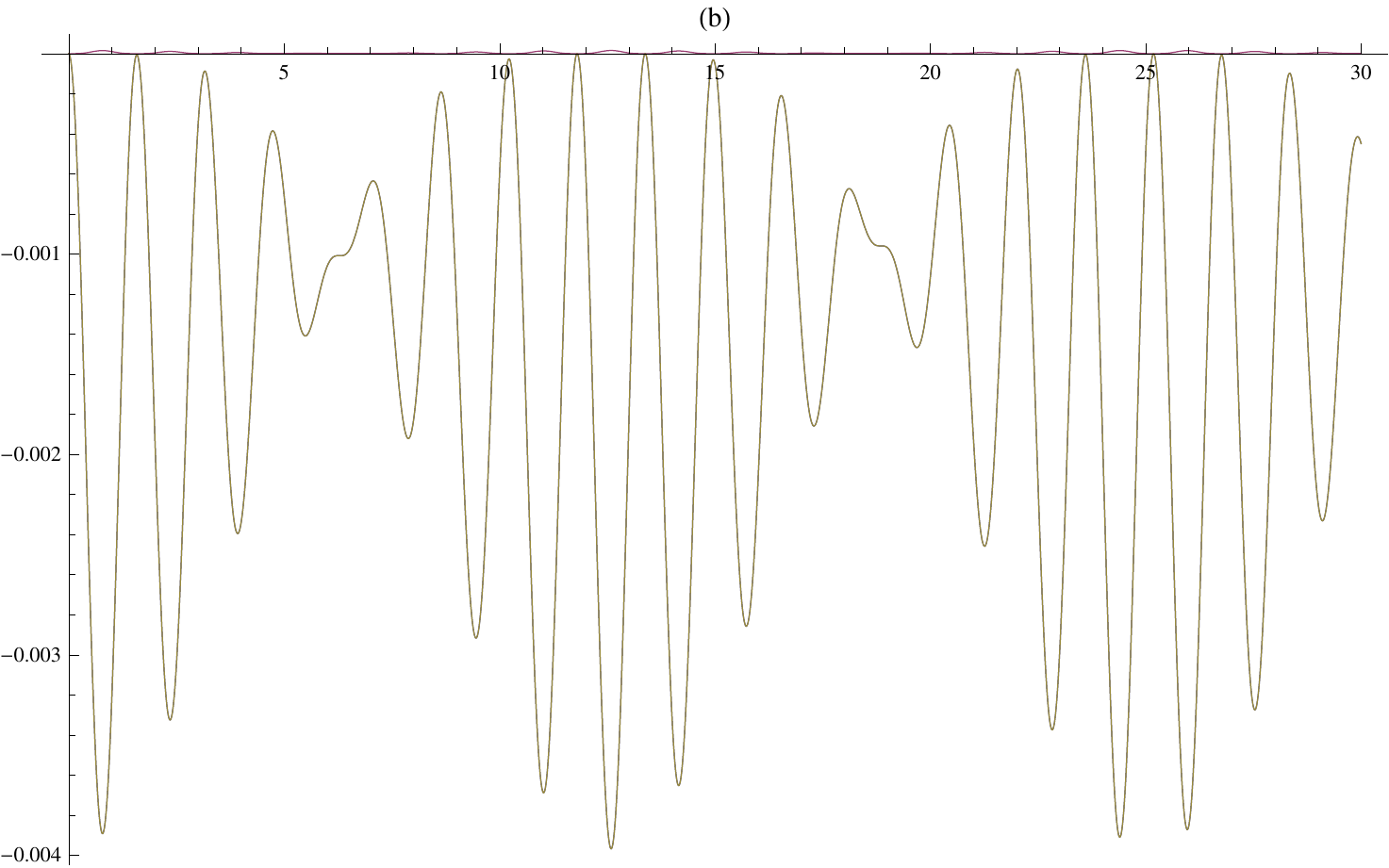}
\caption{Entanglement dynamics for the non-interacting pairwise interaction case. Figure (a) shows the entanglement dynamics of  $\Sigma(X_A,x_{B1},t)$(in blue), $\Sigma(x_{A1},x_{B1},t)$(in purple).  Figure (b) shows the entanglement dynamics of $ \Sigma(X_A,X_B,t)$, $\Sigma(X_A,R_B,t)$(in purple), $\Sigma(R_A,R_B,t)$(in yellow). It can be shown that $\Sigma(X_A,R_A,t)=\Sigma(X_A,R_B,t), \Sigma(X_A,X_B,t)=2\Sigma(X_A,x_{B1},t),$ and $\Sigma(X_A,x_{B1},t)=\Sigma(R_A,x_{B1},t)$ irrespective of the couplings. In these Figures we set $\alpha = \gamma = 0$, $\beta=0.5$,$m=\hbar=1$, $\omega=2$. Under this special choice of parameter $\Sigma(x_{A1},x_{B1},t)$ is almost twice $\Sigma(X_A,x_{B1},t)$ and $ \Sigma(X_A,X_B,t)=\Sigma(R_A,R_B,t)$. These can be easily understood from the definition of COM and relative coordinates. Note that $\Sigma(x_{A1},x_{B1},t) < 0$ for $t > 0$ hence there is always quantum entanglement between $x_1$ and $x_3$ once the coupling between them, $\beta$, was turned on.}
\label{wcCorr1}
\end{figure}

\item  $\beta=\gamma\neq 0$: one-to-all case. Oscillator 1(2) in subsystem A couples to both oscillator 3 and 4 in subsystem B with equal strength.
Fig. 4(a) shows the entanglement dynamics of $\Sigma(X_A,x_{B1},t), \Sigma(x_{A1},x_{B1},t)$ and Fig. 4(b) shows the entanglement dynamics of $ \Sigma(X_A,X_B,t), \Sigma(X_A,R_B,t), \Sigma(R_A,R_B,t)$. Note that $\Sigma(R_A,R_B,t)=0 $ in this case $(\beta=\gamma)$. Hence there is no entanglement among $R_A$ and $R_B$.

\begin{figure}
\includegraphics[width=6cm]{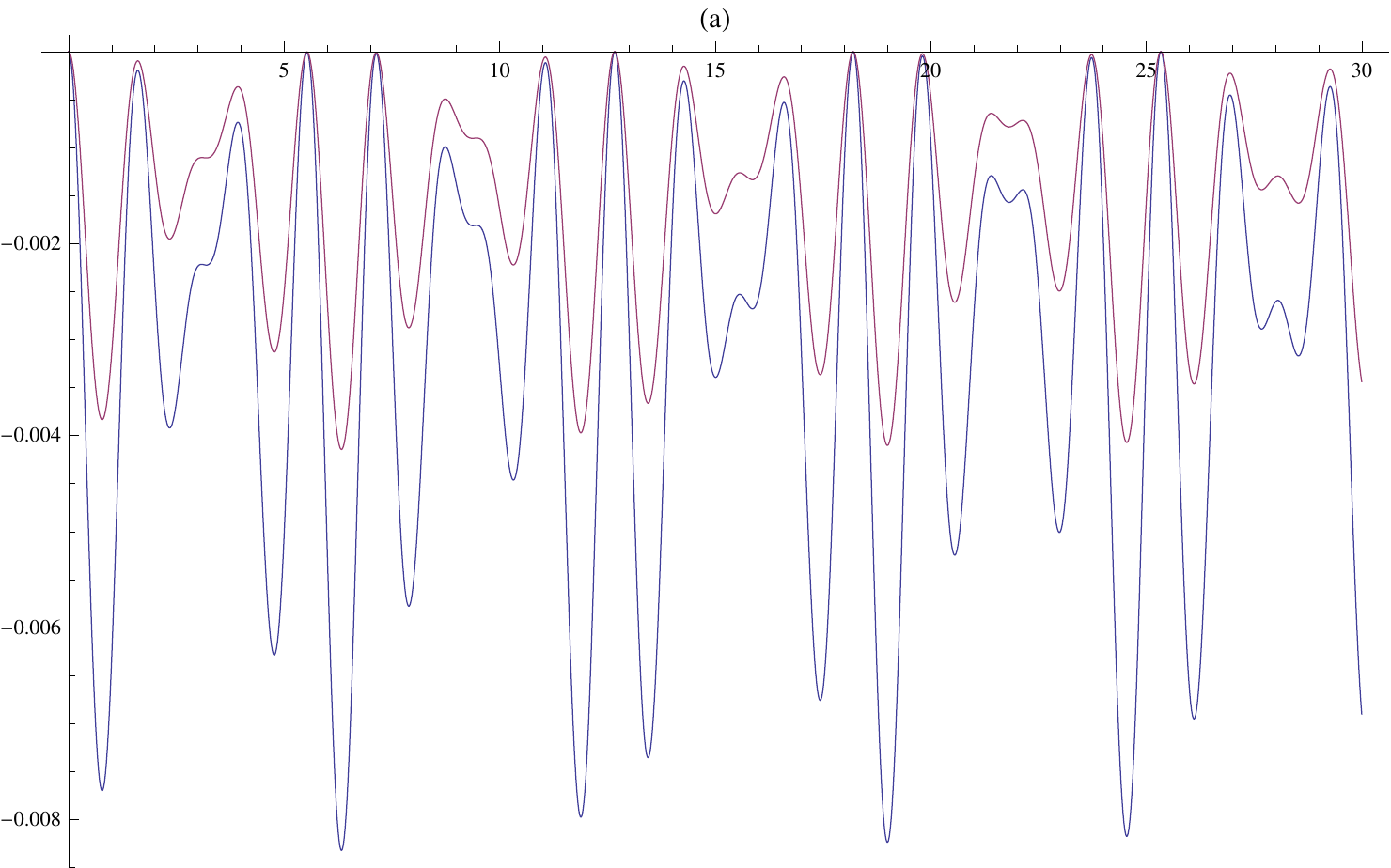}
\includegraphics[width=6cm]{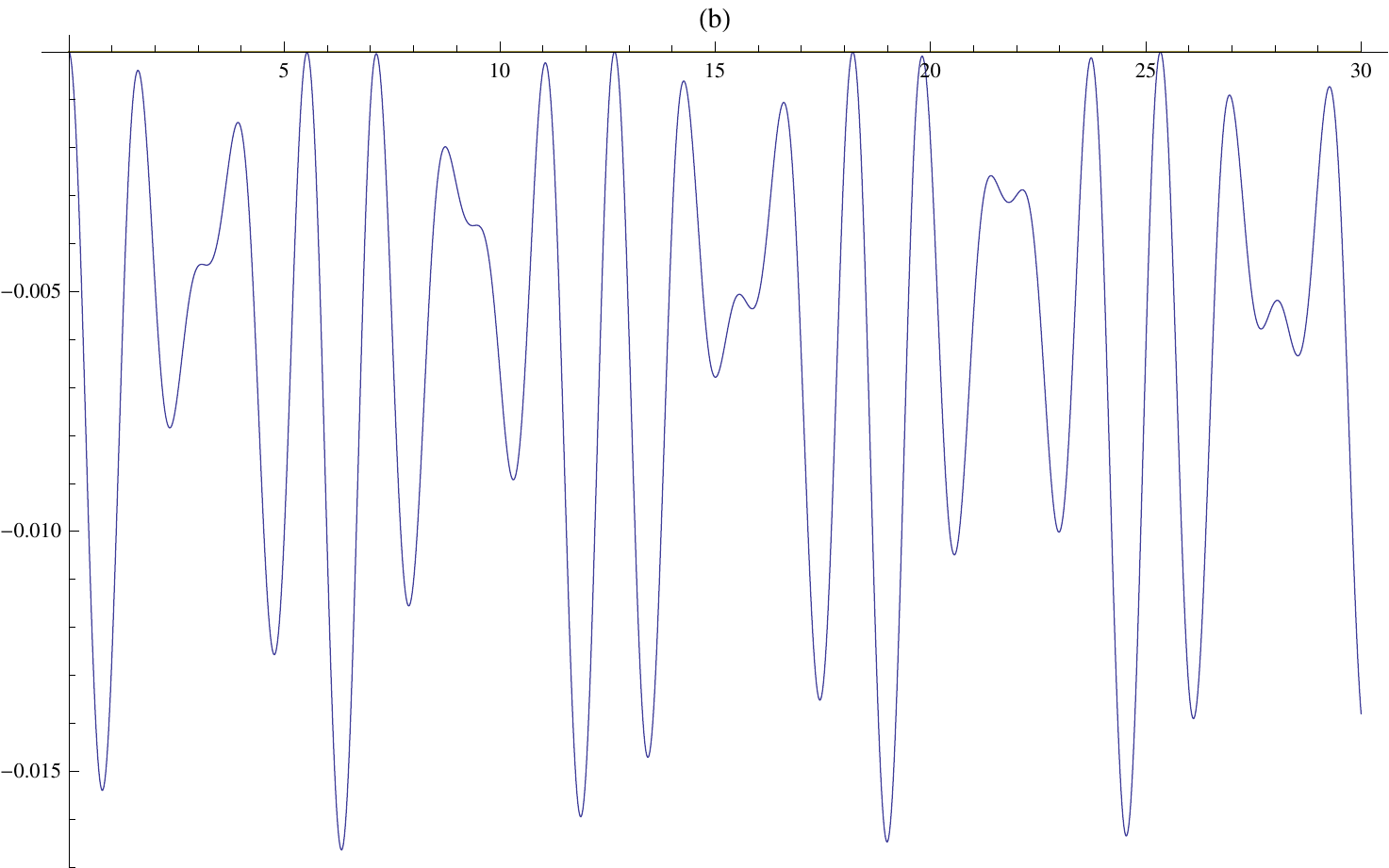}
\caption{Entanglement dynamics for the non-interacting one-to-all case. Figure (a) shows the entanglement dynamics of  $\Sigma(X_A,x_{B1},t)$(in blue), $\Sigma(x_{A1},x_{B1},t)$(in purple). Figure (b) shows the entanglement dynamics of $ \Sigma(X_A,X_B,t)$(in blue). Here $\alpha  = 0$, $\beta= \gamma=0.5$,$m=\hbar=1$, $\omega=2$. Note that under these parameters $\Sigma(X_A,R_B,t)= \Sigma(R_A,R_B,t)=0$ and do not appear in Figure (b). Here one can see for the on-to-all case,  $\Sigma(X_A,x_{B1},t) \approx 2 \Sigma(x_{A1},x_{B1},t)$ hence the COM axion applies as explained in Sec. 3.0.2.}
\label{wcCorr2}
\end{figure}

\end{itemize}

\subsubsection{Entanglement Dynamics for  $\alpha \neq 0$}

Now let's turn on the interactions among the oscillators of each subsystem A, B. In this case $\alpha \neq 0$. We consider the following
subcases.

\begin{itemize}

\item  $\beta \neq 0, \gamma=0$: pairwise interaction case.
Fig. 5(a) shows the entanglement dynamics of $\Sigma(X_A,x_{B1},t), \Sigma(x_{A1},x_{B1},t)$ and Fig. 5(b) shows the entanglement dynamics of $ \Sigma(X_A,X_B,t), \Sigma(X_A,R_B,t), \Sigma(R_A,R_B,t)$. Note that $\Sigma(X_A,R_B,t)=\Sigma(X_A,R_A,t) \geq 0$ in this case and hence there is no entanglement among $X_A$ and $R_B(R_A)$.

\begin{figure}
\includegraphics[width=6cm]{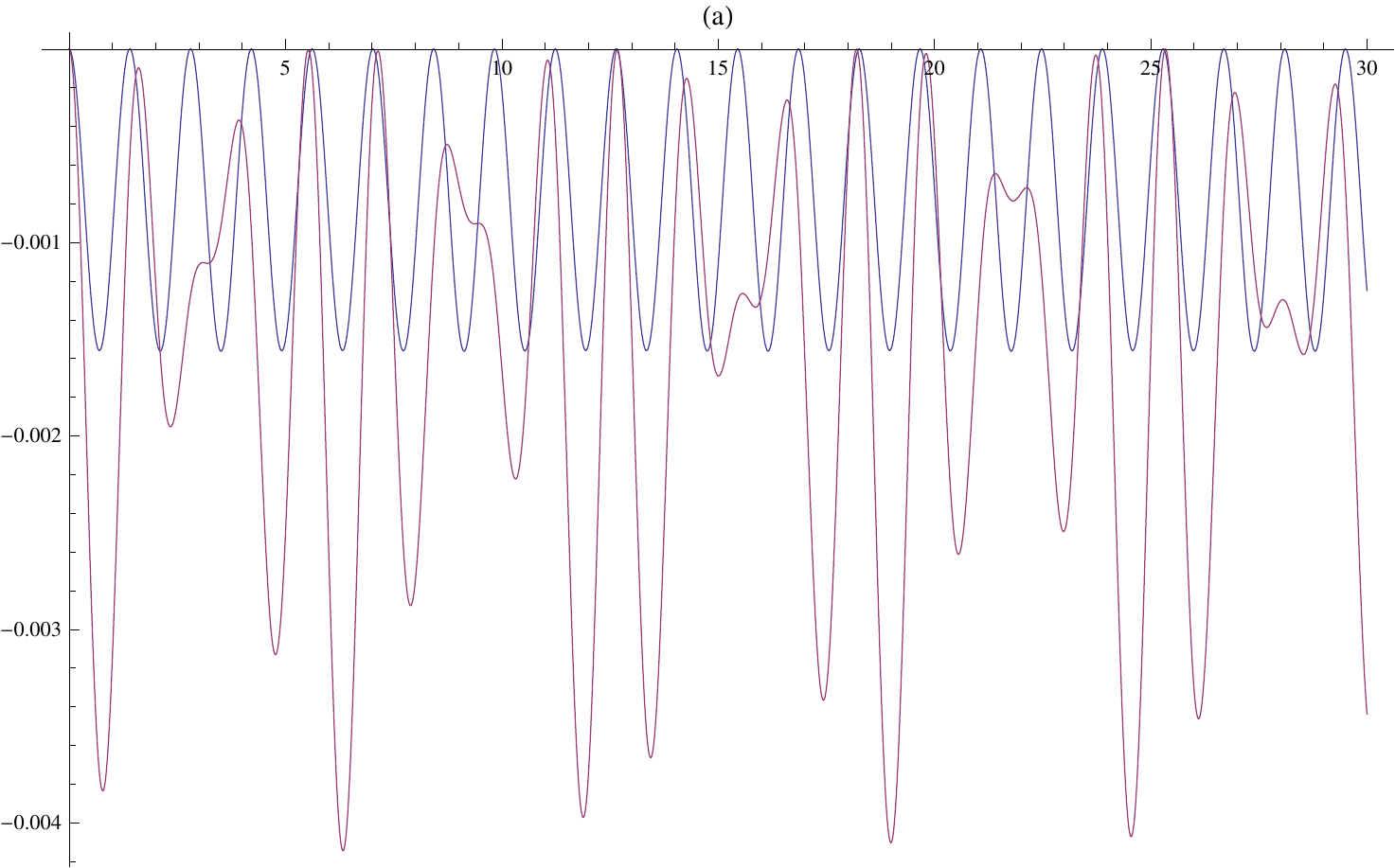}
\includegraphics[width=6cm]{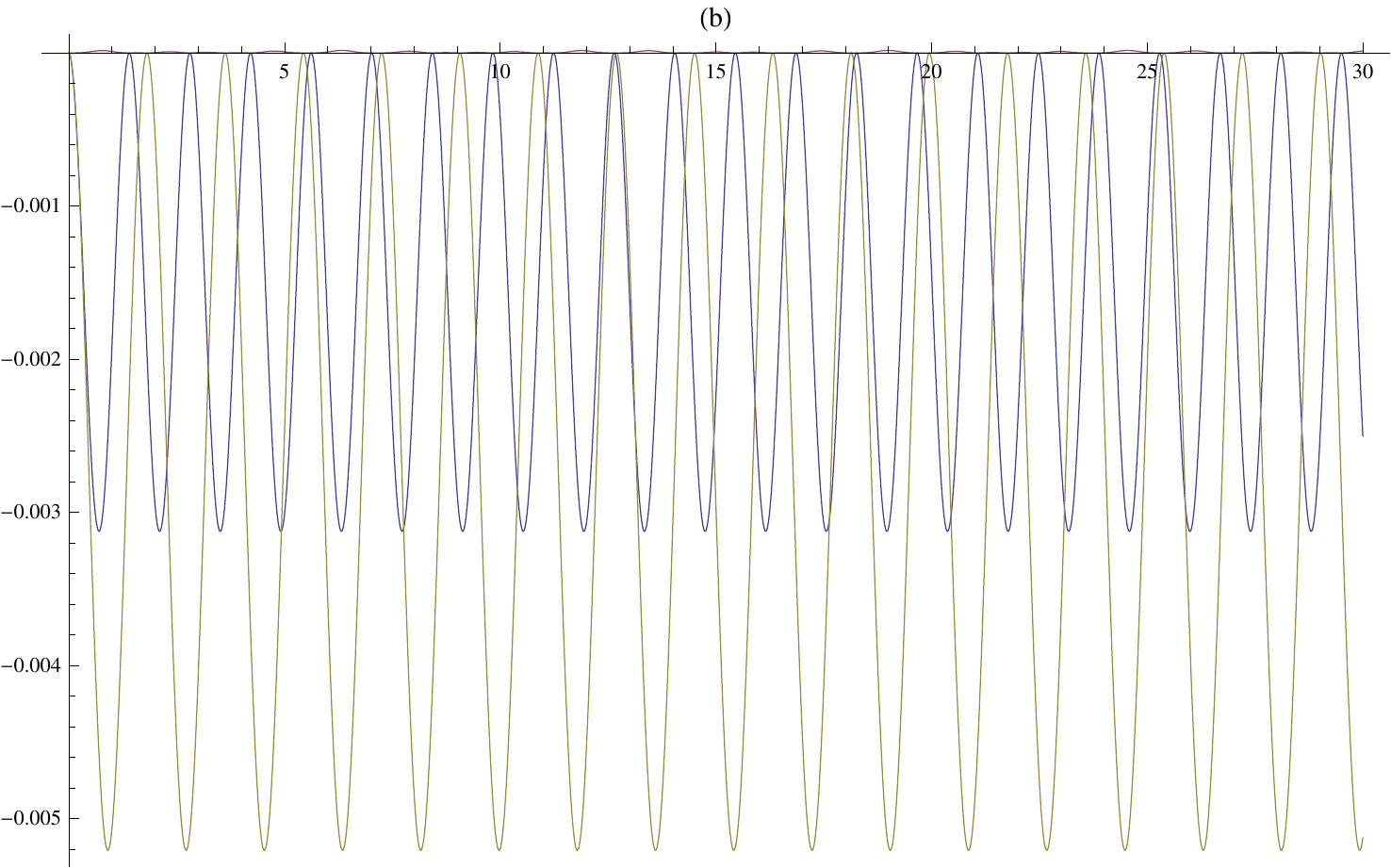}
\caption{Entanglement dynamics for the interacting pairwise interaction case. Figure (a) shows the entanglement dynamics of  $\Sigma(X_A,x_{B1},t)$(in blue), $\Sigma(x_{A1},x_{B1},t)$(in purple). Figure (b) shows the entanglement dynamics of $ \Sigma(X_A,X_B,t)$(in blue), $\Sigma(X_A,R_B,t)$(in purple), $\Sigma(R_A,R_B,t)$(in yellow). Here  $\alpha =\beta=0.5$, $\gamma = 0$,$m=\hbar=1$, $\omega=2$. One can see the effect of the coupling parameter $\alpha\neq 0$ makes the peak values of $\Sigma(X_A,x_{B1},t)$ and $\Sigma(x_{A1},x_{B1},t)$ differ by a sizable amount. $\Sigma(X_A,R_B,t)\geq 0$ (although very small and can hardly read from the figure) in this case, this reflects the separability of the states in the Hilbert spaces of $X_A$ and $R_B$. That means $X_A$ and $R_B$ are always separable. It's interesting to note that the entanglement between $R_A$ and $R_B$ is even stronger than the entanglement among $X_A$ and $X_B$ in this case.}
\label{wcCorr3}
\end{figure}

\item  $\alpha =\beta=\gamma\neq 0$: special one-to-all case.
Fig. 6(a) shows the entanglement dynamics of $\Sigma(X_A,x_{B1},t), \Sigma(x_{A1},x_{B1},t)$ and Fig. 6(b) shows the entanglement dynamics of $ \Sigma(X_A,X_B,t), \Sigma(X_A,R_B,t), \Sigma(R_A,R_B,t)$. Note that in this special case we have $\Sigma(R_A,R_B,t)=0 $ in this case $(\beta=\gamma)$. Hence there is no entanglement among $R_A$ and $R_B$. Also it is found that when $\alpha=\beta=\gamma$, $\Sigma(X_A,X_B,t)=2\Sigma(X_A,x_{B1},t)=4\Sigma(x_{A1},x_{B1},t)$.

\begin{figure}
\includegraphics[width=6cm]{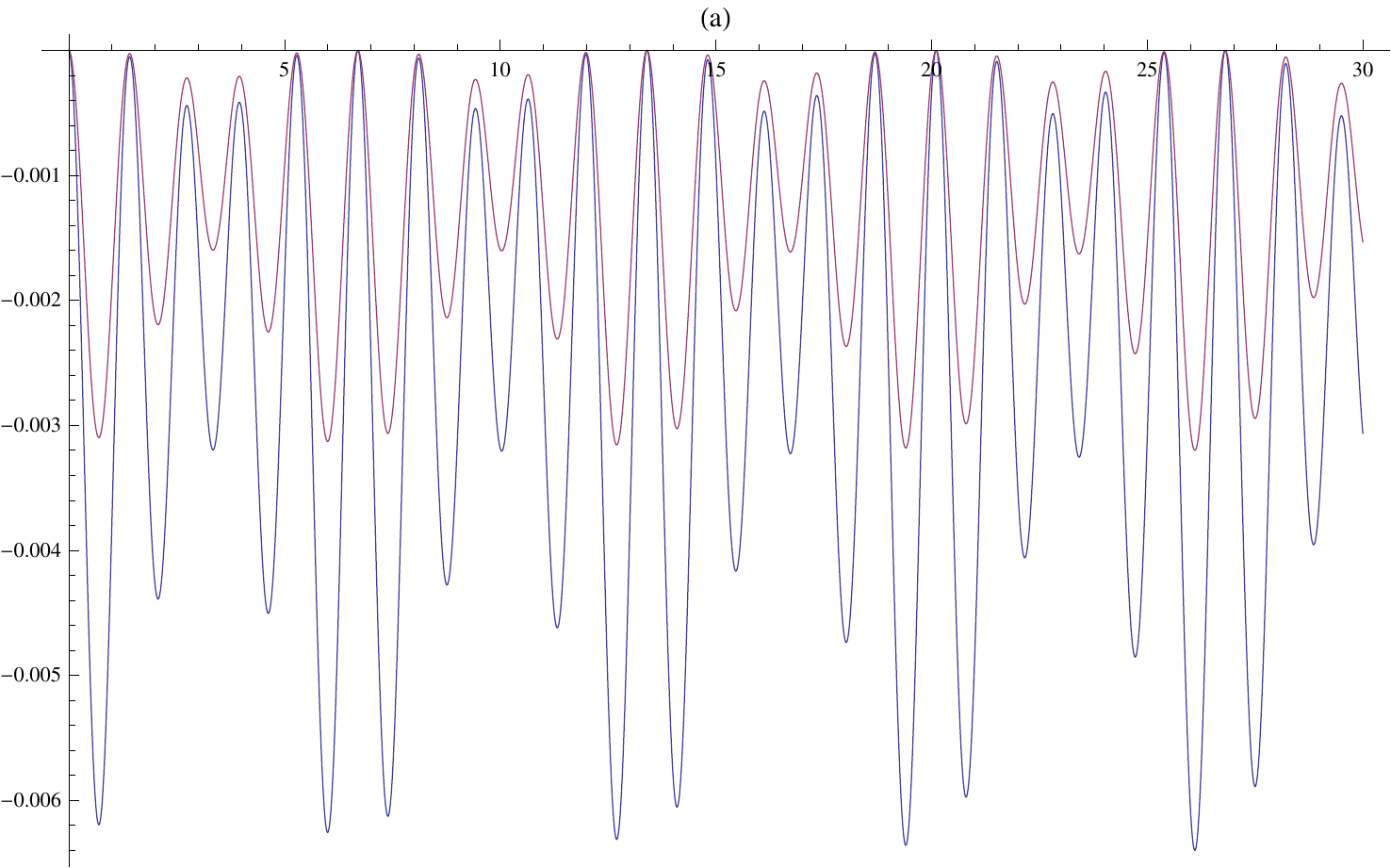}
\includegraphics[width=6cm]{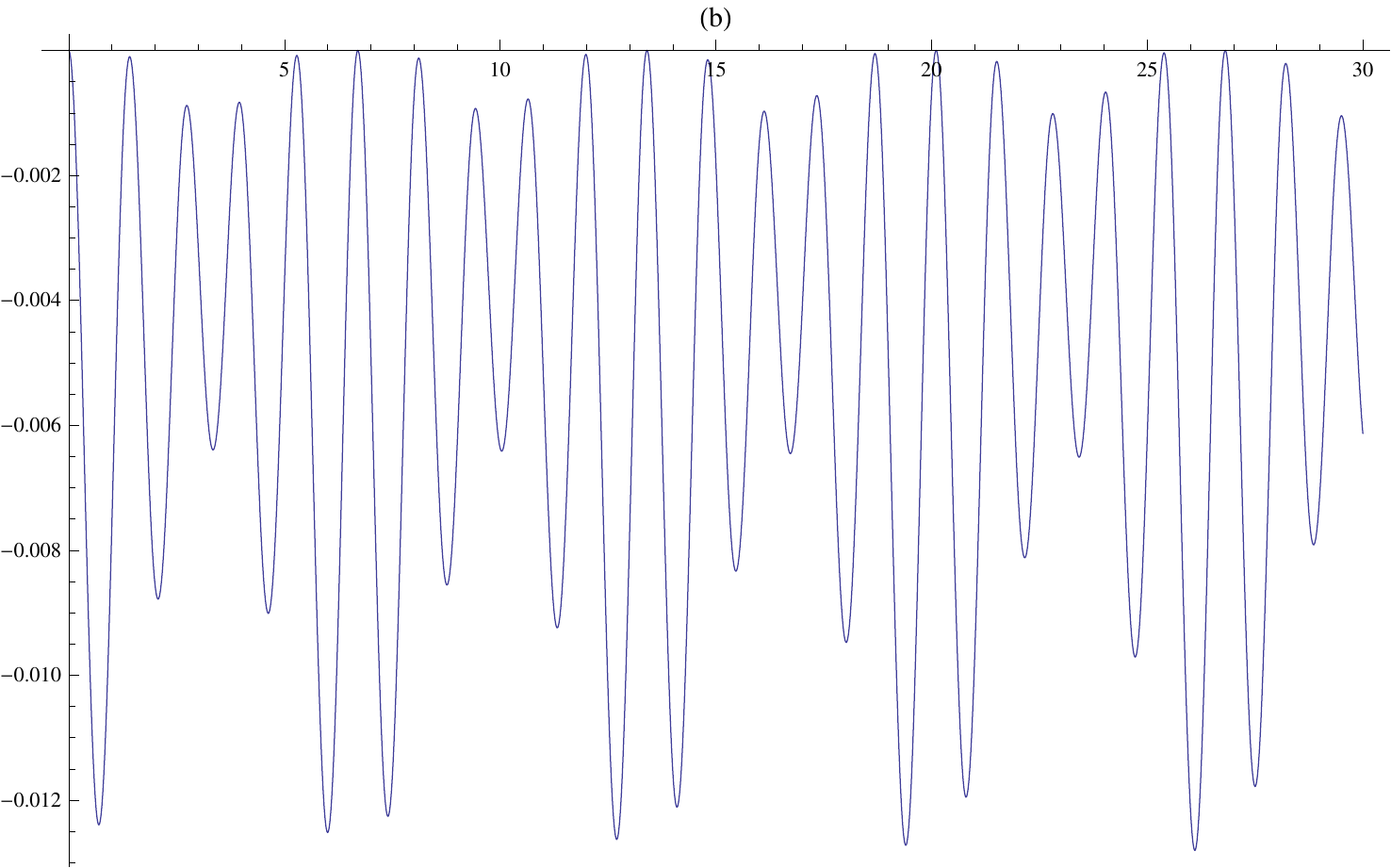}
\caption{Entanglement dynamics for the interacting one-to-all case. Figure (a) shows the entanglement dynamics of  $\Sigma(X_A,x_{B1},t)$(in blue), $\Sigma(x_{A1},x_{B1},t)$(in purple). Figure (b) shows the entanglement dynamics of $ \Sigma(X_A,X_B,t)$(in blue). Here $\alpha  =\beta= \gamma=0.5$,$m=\hbar=1$, $\omega=2$. $\Sigma(X_A,x_{B1},t) = 2 \Sigma(x_{A1},x_{B1},t)$ as expected from the COM axiom even now the coupling between oscillators in each macroscopic object, $\alpha$, is non-zero. Combined with the fact that $ \Sigma(X_A,X_B,t) = 2 \Sigma(X_A,x_{B1},t)$, one has $ \Sigma(X_A,X_B,t) = 4 \Sigma(x_{A1},x_{B1},t)$ under this special $\alpha=\beta=\gamma$ case. Hence the COM coordinate does play a very special role in macroscopic quantum phenomena as illustrated in this section. $\Sigma(X_A,R_B,t)$= $\Sigma(R_A,R_B,t)$=0 and do not appear in Figure (b).}
\label{wcCorr4}
\end{figure}

\item  $\alpha \gg \beta=\gamma\neq 0$: strongly interacting
  one-to-all case. Fig. 7(a) shows the entanglement dynamics of
  $\Sigma(X_A,x_{B1},t), \Sigma(x_{A1},x_{B1},t)$ and Fig. 7(b) shows
  the entanglement dynamics of $ \Sigma(X_A,X_B,t), \Sigma(X_A,R_B,t),
  \Sigma(R_A,R_B,t)$. In this strong coupling regime, the entanglement
  dynamics of $\Sigma(x_{A1},x_{B1},t)$ shows the sudden death and
  revival phenomena. It's interesting to note that
  $\Sigma(X_A,x_{B1},t)\leq 0$ while $\Sigma(x_{A1},x_{B1},t)$ can be
  greater than zero for some finite period of time. This can be easily
  understood as follows: As the coupling $\alpha$
  between the constituents among each macro-objects (A or B) gets
  stronger and stronger, the correlations among them become stronger
  hence the entanglement among the micro-constituents
  $(x_{A1},x_{B1})$ will be strongly affected by the strong coupling
  between $x_{A1}$ and $x_{A2}$ (and also $x_{B1}$ and $x_{B2}$). This
  will make the entanglement between $x_{A1}$ and $x_{B1}$ vanish for
  some time intervals. However the two macro-objects (A and B) will interact mainly through the coupling $\beta$ in the one-to-all case  hence the entanglement between $X_{A}$ and $x_{B1}$ still holds.

\begin{figure}
\includegraphics[width=6cm]{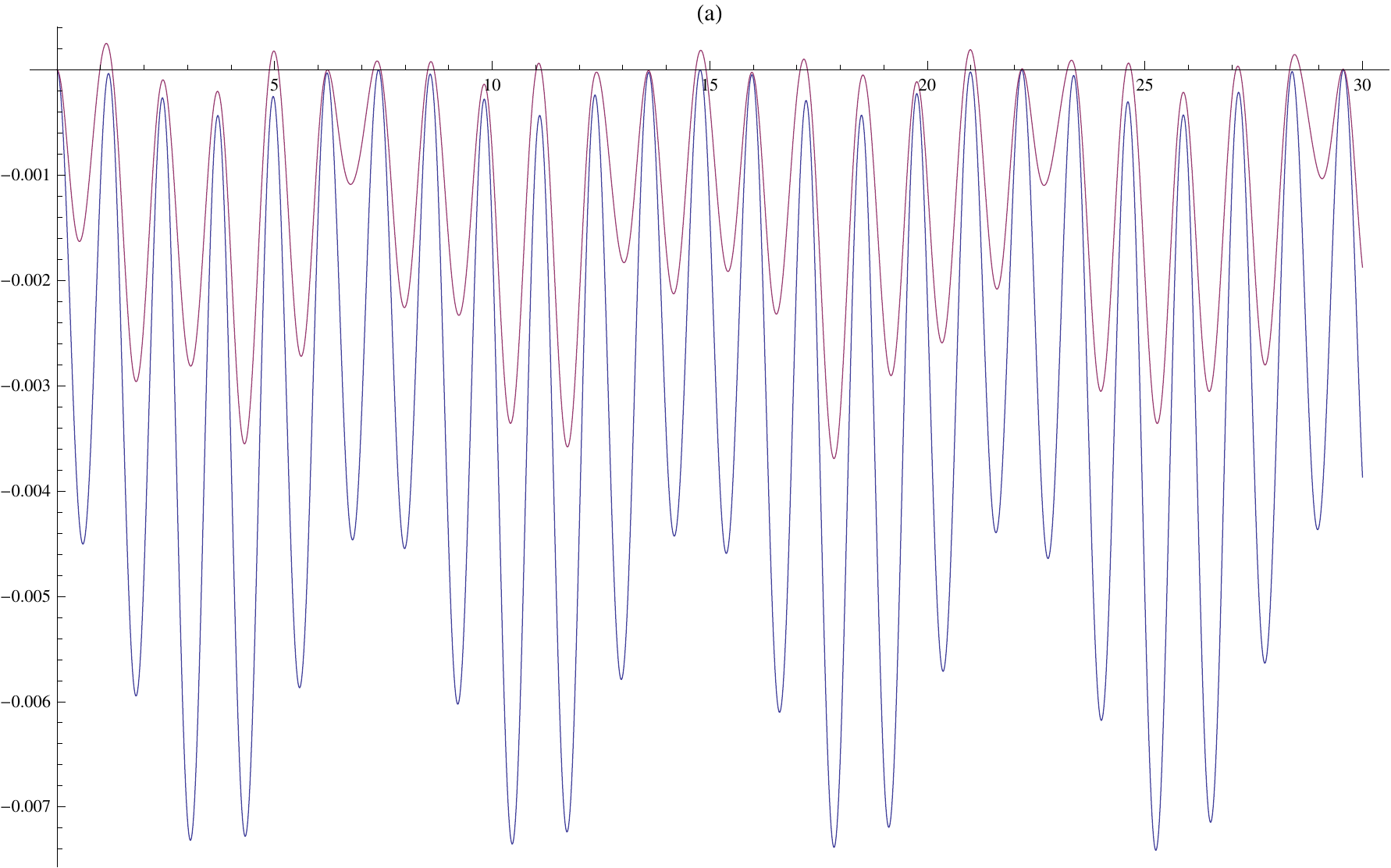}
\includegraphics[width=6cm]{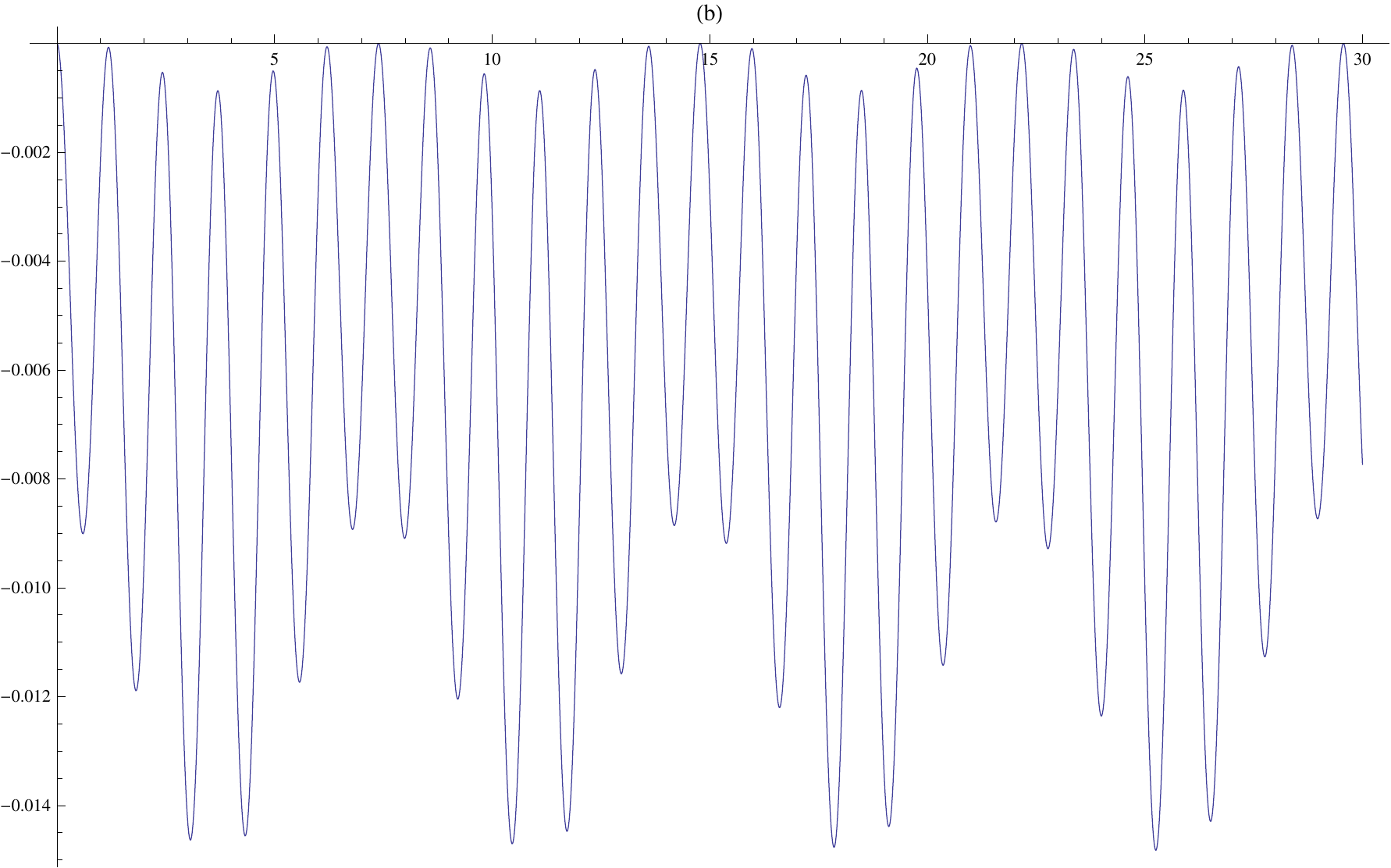}
\caption{Entanglement dynamics for the strong interacting one-to-all case. Figure (a) shows the entanglement dynamics of  $\Sigma(X_A,x_{B1},t)$(in blue), $\Sigma(x_{A1},x_{B1},t)$(in purple). Figure (b) shows the entanglement dynamics of $ \Sigma(X_A,X_B,t)$(in blue). Here $\alpha=1.5,  \beta= \gamma=0.5$,$m=\hbar=1$, $\omega=2$. Two major feature differences as compared to $\alpha=\beta=\gamma$ case. One is that $|\Sigma(X_A,x_{B1},t)|$ is larger than twice the amount of  $|\Sigma(x_{A1},x_{B1},t)|$. The other feature is the entanglement dynamics of $\Sigma(x_{A1},x_{B1},t)$ shows the sudden death and revival phenomena while those for $\Sigma(X_A,x_{B1},t)$ does not. $\Sigma(X_A,R_B,t)$= $\Sigma(R_A,R_B,t)$=0 and do not appear in Figure (b).}
\label{wcCorr5}
\end{figure}

\item  $  \beta=\gamma \gg \alpha \neq 0$: weak interacting one-to-all case.  Fig. 8(a) shows the entanglement dynamics of $\Sigma(X_A,x_{B1},t), \Sigma(x_{A1},x_{B1},t)$ and Fig. 8(b) shows the entanglement dynamics of $ \Sigma(X_A,X_B,t), \Sigma(X_A,R_B,t), \Sigma(R_A,R_B,t)$. This is similar to the $\alpha=0, \beta=\gamma$ case. The effect of the small nonzero coupling $\alpha$ will slightly reduce the amount of entanglement.

\begin{figure}
\includegraphics[width=6cm]{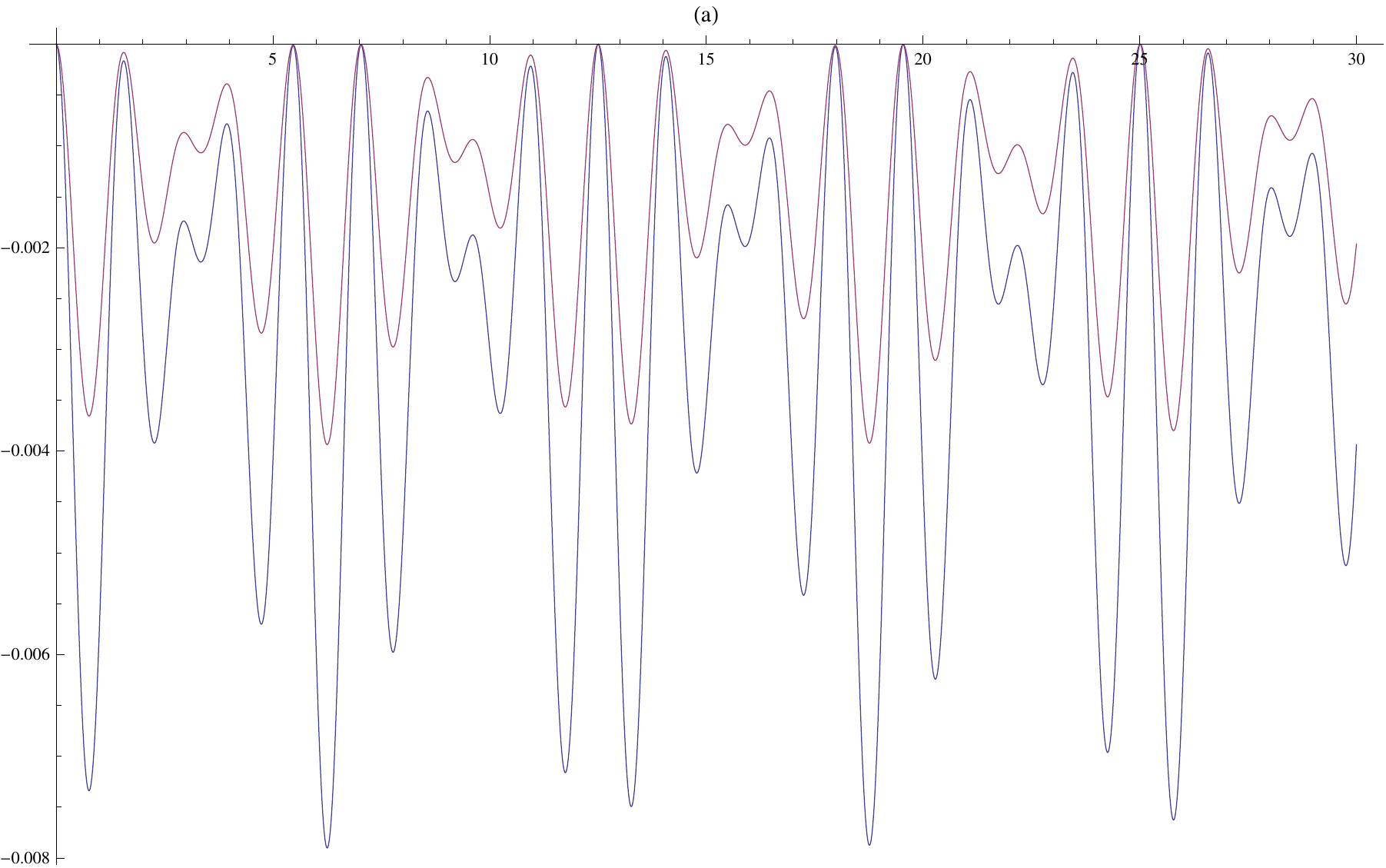}
\includegraphics[width=6cm]{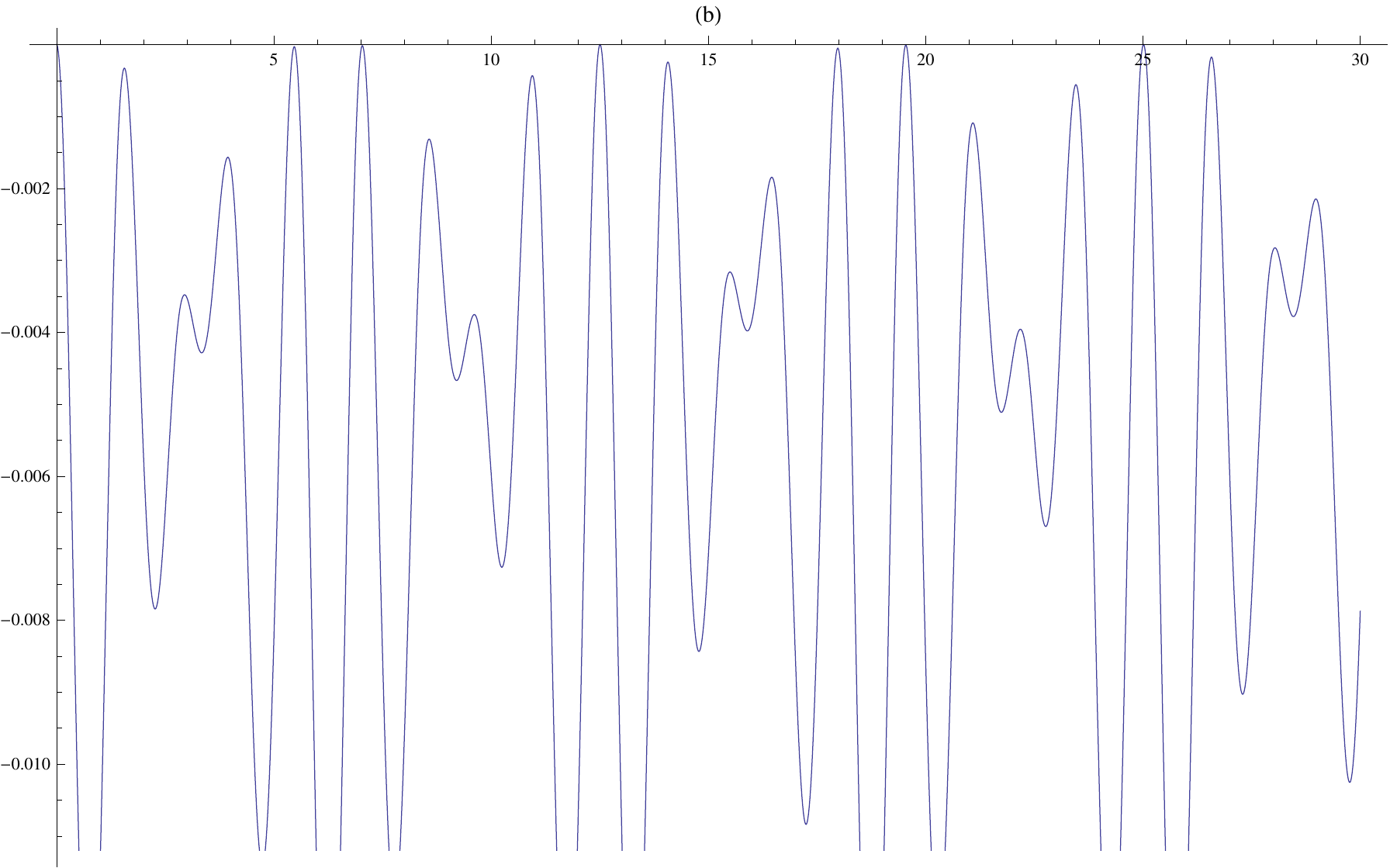}
\caption{Entanglement dynamics for the weak interacting one-to-all case. Figure (a) shows the entanglement dynamics of  $\Sigma(X_A,x_{B1},t)$(in blue), $\Sigma(x_{A1},x_{B1},t)$(in purple). Figure (b) shows the entanglement dynamics of $ \Sigma(X_A,X_B,t)$(in blue). Here $\alpha=0.1,  \beta= \gamma=0.5$,$m=\hbar=1$, $\omega=2$. Note that under these parameters $\Sigma(X_A,R_B,t)= \Sigma(R_A,R_B,t)=0$ and do not appear in Figure (b). Here one can see for the on-to-all case,  $\Sigma(X_A,x_{B1},t) \approx 2 \Sigma(x_{A1},x_{B1},t)$ hence the COM axiom applies as explained in Sec. 3.0.2. The effect of the coupling makes the amount of entanglement being slightly reduced as compared to the noninteracting case as those in Figure 3(where $\alpha=0$). }
\label{wcCorr6}
\end{figure}

\end{itemize}

In summary, the 4 harmonic oscillators case gives a concrete example
supporting the physical arguments and analysis in previous sections. Here we provide the
entanglement structure and dynamics depending on whether there is interaction (coupling strength =
$\alpha$) among constituent oscillators of each subsystem. We analyze two special
cases: one-to-one(pairwise) and one-to-all in detail.
(i). No quantum entanglement between the center-of-mass  $X_A$ and the  relative coordinates $R_A$($R_B$).
(ii). For non-interacting $(\alpha=0)$ one-to-one case $\Sigma(X_A,x_{B1},t) \approx 0.5\Sigma(x_{A1},x_{B1},t)$
as can be easily understood from the definition of the center-of-mass coordinate. (Fig.3)
(iii). For non-interacting $(\alpha=0)$ one-to-all case $\Sigma(X_A,x_{B1},t) \approx 2\Sigma(x_{A1},x_{B1},t)$ hence supporting the COM axiom. (Fig.4)
(iv). For interacting $(\alpha >0)$ one-to-one case one can see the effect of the coupling on the entanglement dynamics.(Fig.5)
(v). For special interacting one-to-all case $(\alpha=\beta=\gamma >0)$ $\Sigma(X_A,X_B,t)=2\Sigma(X_A,x_{B1},t)=4\Sigma(x_{A1},x_{B1},t)$.
For this special case, the model processes  a larger symmetry and hence the clear-cut behavior of the
corresponding entanglement dynamics. This also supports the analysis in the previous section. (Fig.7)
(vi). In the one-to-all cases,  as the coupling $\alpha$ changes from weak to strong (compared with $\beta$), we see how the strong coupling alters the entanglement dynamics and drives the sudden death and revival of entanglement among constituents between different subsystems. (Fig.7,8)

In short, the 4 harmonic oscillators case gives a concrete example illustrating the physical arguments and analysis in previous sections, lending concrete support to the COM axiom.

\section{Discussions}

In this paper we try to understand MQP from the quantum entanglement perspective by asking several key questions and suggesting some pathways to understand them. As a background to this question we have in two earlier papers \cite{MQP1,MQP2} asked the question, what is meant by macroscopic? -- Is it the size of an object? Its contents, as measured by the number of particles or components? What about its composition? Here the coupling strength of its constituents matter. What about its `appearance'? -- in the sense that the energy scale or precision level one probes the object matters, etc. Against this backdrop,  we ask these questions:

1) Knowing that macroscopic objects are made up of micro-constituents, what does it mean when one refers to ``the  entanglement between two macroscopic objects"?

2) How does the entanglement between micro-constituents contribute to entanglement in the macroscopic objects?

3) How to quantify macroscopic entanglement based on the entanglement structure and dynamics of its constituents?

The partial answers we provided are, respectively, as follows:

1) One really needs to start with the entanglement between the
    constituent particles.  But immediately one encounters the puzzle
   as to what would be regarded as constituent, since each constituent
   particle is made up of  sub-constituent particles. Here unavoidably
   a) \textit{the level of structure scheme} of macro or even meso
   objects enters in an essential way. When we talk about the
   entanglement between two objects are we referring to the
   entanglement between the quarks of one object with that of the
   other, or the molecules of each object? The obvious difference
   between them is that quarks are strongly coupled and the interactions among them is short-ranged while interactions among molecules
   are much weaker but with longer effective interaction range.` Thus b) the \textit{coupling strength of
   constituents} should enter.  Thirdly, c) Assuming that one can find a measure of entanglement between two macro objects of a definite value, call it $E(M)$,  and we know that their constituents fall under levels of structures, how does one derive $E(M)$ from the entanglements between these micro-constituents, call them $E(\mu)$. This problem, let's call it \textit{the ``sum-rule" question}, is highly nontrivial. The quarks and gluons are strongly coupled and are expected to have stronger entanglements, but their interaction is short ranged, thus their contributions to E(M) may not be as high as the van der Walls force between molecules. A measure of entanglement at the macro-level could involve some coarse-graining of the contributions of the sub-constituents. But what is the criteria for coarse-graining? If there is some way to express $E(M)$ in terms of $E(\mu)$ would the sum-rule include a weighing factor attributed to the entanglement measure from each level of structure? Is there some kind
   of hierarchical ordering in systemizing the weighing factors  in the contributions from each level of structure for $E(\mu)$ to add up to a finite value? These are interesting questions worthy of further investigation. We have only some glimpses of an answer to this question: They don't usually add up, but depend on the coupling pattern of the constituents and what and how different parties are entangled.
   This last point is what enables one to address questions 2) and 3) below.

2) Acknowledging the complexity of what makes up a macroscopic object,
   constructing or \textit{identifying a set of collective variables}
   from the constituents in each level of structure is therefore an
   essential step towards organizing their contributions.  We showed
   the conditions under which the center of mass of an object plays a special role in MQP. Its distinguished role is highlighted by the fact that the CoM variables are in general decoupled from the relative variables
and the dynamics of the CoM variables behaves differently from the relative variables.

To understand how the \textit{coupling pattern amongst the constituents} of the two macro objects enters into the picture, we consider two types of coupling, the 1-to-1 versus the 1-to-all. For the 1-1 or pairwise interactions with equal strength, the entanglement is independent of the size $N$ of the macroscopic objects. The 1-1 interaction pattern is the only one for which the CoM and relative coordinate couplings are effectively the same. In the 1-to-all case the relative coordinates are decoupled and the CoM coupling scales with N. In this case we expect the entanglement between the CoM variables to increase with increasing size of the macroscopic objects and survive at higher temperatures.
Note that this scaling property is proven here for Gaussian systems only.
We have also considered Gaussian fluctuations around the 1-to-1 and
1-to-all patterns and shown that these conclusions continue to
hold in the presence of such perturbations in the thermodynamic limit.

Our detailed analysis of the specific model, that between
two pairs of constituent oscillators [(3) below] also displays further distinct behavior of the CoM in terms of entanglement dynamics, like the absence of sudden death of entanglement in the 1-to-all case, etc.

3) We pause here to compare our findings with prior work on
   entanglement structure of Gaussian systems such as that carried out
   systematically by Adesso and Illuminati
   \cite{Illuminati2005,ConTangle,AdeIll2007}. In our analysis we have shown that the one-to-all interaction pattern, when viewed in the transformed coordinates, corresponds to only CoM-CoM interactions being non-vanishing, which moreover scale linearly with $N$. This ability to concentrate the interactions between two macro object each consisting of $N$ oscillators into interactions between $2$ modes only is related to the results of \cite{Illuminati2005}. In that paper it is shown that if a Gaussian state of $N+N=2N$ oscillators is bi-symmetric, then there exists a unitary transformation which acts on each object separately and results in $2N-2$ uncorrelated modes and $2$ correlated modes, one in each object. We showed through the coupling pattern analysis (in Sec. 3) and explicit results of a 4-oscillator system (in Sec. 4) that these two correlated modes correspond to the CoM variables which serve to capture the essence of the interactions between the two macroscopic objects completely.

Reference \cite{Illuminati2005} has shown that the $1\times 1$ entanglement vanishes in the limit $N\rightarrow \infty$, whereas the $N\times N$ entanglement diverges. This shows that 1x1 entanglement (even if summed over all pairs) is not enough to account for the macroscopic entanglement between two objects. The localized entanglement, in our case between the CoM variables, quantifies the total entanglement between the two objects, and its divergence with N is consistent with our result that the interaction strength is proportional to N.

In contrast to the one-to-all case, it can be shown that in the pairwise coupling case no coordinate transformation can bring forth a "concentration" of the interactions between the macro objects. All transformed coordinates are still pairwise coupled with equal coupling strengths. This does not contradict the results of \cite{Illuminati2005} however, since the pairwise coupling scheme is not bisymmetric. As a result we can not apply the general results about bisymmetric Gaussian states to the pairwise coupling case. In this case the CoM fails to capture the interactions between the two objects completely.

4) Before drawing implications of these results we should  make two points clear here:  A)  Reference \cite{Illuminati2005} considered states (equivalently, covariance matrices) whereas we analysed Hamiltonians, which are more directly related to the dynamics. However, when considering thermal states or dynamics following initially uncorrelated states, the properties of evolved states follow closely the properties of the Hamiltonian. Because of this our results can be related to theirs.
B) Our results in Sec. 3 addresses how the coupling strength depends on the coupling pattern and varies with the constituent versus the collective variables. Statements made stemming from there pertaining to entanglement strength assume that entanglement is proportional to coupling strength. Even though this is intuitively acceptable, the relation between entanglement magnitude and coupling strength remains to be made more explicit.

Allowing for this connection we may continue our discussions further: The fact that the entanglement between two (or more) macroscopic objects coupled in the \ota pattern grows as $N$ increases (in some power or pattern) is easy to understand:  
Since the \ota coupling is identical for all constituent particles, it does not matter which object the particles are the constituents of, the partition between the two objects can be viewed as a conceptual artifact.  It can as well be regarded as the entanglement between two parts of one macro object. One can also turn
this around and introduce the entanglement between the (CoMs of the)
two halves as providing a measure of the \textit{entanglement strength
of a macroscopic object.}

Furthermore, since in reality the coupling strength often depends on
the separation between the constituent particles the case studied here
is realized only in short-ranged yet strongly-interacting cases, such
as nucleons in a nucleus, whose interaction is quite adequately
depicted by a square well potential. 
Note that it is the one-to-all pattern, rather than the strength, of nuclear interactions that allows us to use them as an example here.
Motivated by this, even apart from MQP considerations the way how entanglement varies with coupling strength and pattern can in principle be used to explore the \textit{entanglement entropy for a quantum many-body system}. 
E.g., depending on the coupling pattern the entanglement entropy can scale differently with the number of particles in the system, from which one can see how the proportionality shifts, say, from the volume to the surface or perhaps even somewhere in between.
This carries a different meaning from the conventional one where for
short-ranged forces the entanglement entropy of a quantum many-body
system is related to the surface area of an artificially introduced
partition. (For a review on this subject see, e.g., \cite{Campo}. The
original area theorem found by Bekenstein for black holes is in a more
natural setting because of the existence of event horizons.)  The analysis of the \ota model here brings out a different facet of the problem, namely, the coupling strength and pattern of constituents, in defining a suitable measure of entanglement entropy for quantum many-body systems.  

5) To quantify these behavior, specifically aiming at how entanglement
   depends on how the constituents are organized by their coupling
   strengths and how the CoM axiom applies to the entanglement of
   macro-objects, we analyzed the entanglement structure of a four
   oscillator system in detail, not only in the kinematics as is done
   in the two interaction patterns and in the proofs, but here also in
   the dynamics of entanglement.  By assigning different strengths to
   the three pairwise coupling constants we can investigate the
  { entanglement differences between the intra- and
   inter- level of structures.}  By setting the coupling strength
   between the intra-level constituents to be weak, we reproduce the
   results for the 1-to-1 pattern, while keeping the three coupling
   strengths the same we obtain the results for the 1-to-all pattern.
    We verify the dominance in the entanglement of the CoM variables by showing the existence of sudden death in the entanglement dynamics of inter-level constituents and the persistence of the entanglement between the CoM variables. This model continues what was studied before in the entanglement dynamics of say two two-level atoms interacting with common field modes \cite{ASH,Kanu} or two oscillators interacting with a common field \cite{CYH08,RecPaz}, but with a new angle toward issues in MQP. In terms of practical relevance, we commented on the 1-to-all pattern above, as relating to the entanglement entropy of a many-body system. For the 1-to-1 pattern it can be applied to finding the entanglement in materials with layered structures, from two BEC slabs, to graphene, even to the two strands of DNA {sequences where the DNA A(C) couples pairwise with DNA T(G).} Our effort here is to search for pathways into analyzing the entanglement structure and dynamics of macro-objects. The field is wide open for various theoretical probings -- starting even with asking good questions -- with  intellectual rewards, and the  results will bear immediate practical applications.

\noindent{\bf Acknowledgment} We wish to thank Dr. J. T. Hsiang  for
sharing his calculation results for the entanglement pattern of
three coupled oscillator systems; and R. Zhou and K. Sinha for discussions. CHC
thanks the support from National Center for Theoretical Science(South)
of Taiwan. This work is supported by the National Science Council of
Taiwan under grant NSC 101-2112-M-006-003. YS was a Visiting Graduate
Fellow at Perimeter Institute in the Spring semester of 2013. BLH thanks Professor M. C. Chu of the Chinese University of Hong Kong and Professors Y. S. Wu and J. Q. You of Fudan University for their kind hospitality during his visits in the Spring of 2013 when this work was carried out.

\end{document}